\documentclass[useAMS,usenatbib]{mn2e}
\usepackage{epsfig}
\usepackage{aas_macros}
\usepackage{fixltx2e}
\title[Gamma-ray Attenuation and the UV Background]{GeV Gamma-ray Attenuation and the High-Redshift UV Background}
\author[R. C. Gilmore et al.]{Rudy C. Gilmore$^{1}$\thanks{E-mail:rgilmore@physics.ucsc.edu}, Piero Madau$^{2}$, Joel R. Primack$^{1}$, Rachel S. Somerville$^{3}$, \newauthor Francesco Haardt$^{4}$\\
$^{1}$Department of Physics, University of California, Santa Cruz, CA 95064 \\
$^{2}$Department of Astronomy \& Astrophysics, University of California, Santa Cruz, CA 95064 \\
$^{3}$Space Telescope Science Institute, 3700 San Martin Dr., Baltimore, MD 21218 \\
$^{4}$Dipartimento di Fisica e Matematica, Universit\'{a} dell'Insubria, Via Valleggio 11, 22100 Como, Italy}
\begin{document}
\date{\today}

\newcommand{\plottwo}[2]
           {\begin{center} \leavevmode \psfig{file=#1,width=8.5cm,clip=}
                            \hfill \psfig{file=#2,width=8.5cm,clip=} \end{center}}
\def\lesssim{\lower.5ex\hbox{$\; \buildrel < \over \sim \;$}}
\def\gtrsim{\lower.5ex\hbox{$\; \buildrel > \over \sim \;$}}
\newcommand{\feschi}{\mbox{${f}_{\rm esc, HI}$}}

\pagerange{\pageref{firstpage}--\pageref{lastpage}} \pubyear{2009}

\maketitle
\label{firstpage}

\begin{abstract}

We present new calculations of the evolving UV background out to the
epoch of cosmological reionization and make predictions for the amount
of GeV gamma-ray attenuation by electron-positron pair production.
Our results are based on recent semi-analytic models of galaxy
formation, which provide predictions of the dust-extinguished UV
radiation field due to starlight, and empirical estimates of the
contribution due to quasars. We account for the reprocessing of
ionizing photons by the intergalactic medium. We test whether our
models can reproduce estimates of the ionizing background at high
redshift from flux decrement analysis and proximity effect
measurements from quasar spectra, and identify a range of models that
can satisfy these constraints. Pair-production against soft diffuse
photons leads to a spectral cutoff feature for gamma rays observed
between 10 and 100 GeV.  This cutoff varies with redshift and the
assumed star formation and quasar evolution models.  We find only
negligible amounts of absorption for gamma rays observed below 10 GeV
for any emission redshift.  With observations of high-redshift sources
in sufficient numbers by the {\it Fermi Gamma-ray Space Telescope} and
new ground-based instruments it should be possible to constrain the
extragalactic background light in the UV and optical portion of the
spectrum.
\end{abstract}


\section{Introduction}
Interactions between photons via electron-positron pair production can
have a substantial effect on the observed spectra of extragalactic
gamma-ray sources.  This process, which can occur when the required
threshold energy of twice the electron mass is present in the
centre-of-mass frame, removes gamma rays en route to the observer
\citep{gould67}, and provides a link between high-energy astrophysics
and the extragalactic background light (EBL) -- the integrated
luminosity of the universe at UV, optical, and IR frequencies.

Attempts to measure the present-day EBL through absolute photometry in
the optical and IR portion of the spectrum are hampered by the bright
foregrounds of the Milky Way and zodiacal light from dust in the solar
system, as well as calibration uncertainties
\citep{hauser&dwek01}. Measurements in the near- and far-IR are
available from the DIRBE and FIRAS experiment on the {\it COBE}
satellite (\citealp{wright&reese00,wright01,cambresy01};
\citealp*{levenson07};
\citealp{hauser98,lagache00,wright04,fixsen98}).  A tentative
determination of the optical EBL was made in
\citet*{bernstein02a,bernstein02b} with revision in
\citet{bernstein07}.  Number counts of observed galaxies provide a
robust lower limit to the EBL, but the degree to which these
measurements converge can be controversial.  The analysis of the
integrated galaxy counts in seven optical and near-IR bands by
\citet{madau00} showed that flux from direct starlight converged to a
fairly low level, below that claimed in the several DIRBE detections
of near-IR flux.  Recent experiments such as the {\it Galaxy Evolution
  Explorer} ({\it GALEX}) satellite, {\it ISO}, and {\it Spitzer} have
provided counts data in a variety of non-optical wavelengths, placing
lower limits on the local EBL at most wavelengths of interest.  A more
thorough discussion of current EBL constraints can be found in
\citet*{gilmoreEBL} and \citet*{primack08}.

A number of techniques have been employed to build cosmological models
of the emission of light from galaxies at UV to far-IR wavelengths.
These include interpolation and extrapolation of cosmological
observables such as star formation rate (\citealp*{kneiske02};
\citealp{kneiske04}; \citealp*{razzaque09}), and luminosity functions
\citep*{franceschini08}. Another method evolves backward in time the
local galaxy population, usually by assuming that the luminosity
density changes with redshift as a power law at all wavelengths
\citep*{stecker06}.  A third class of techniques models the galaxy
population forward in time, beginning with cosmological initial
conditions (\citealp{primack01}; \citealp*{primack05}).

Our goal in this paper is to build a suite of models of the evolving
background light produced by stars and quasars, with a focus on the
optical-UV background out to high redshift.  A preliminary report on
this work appeared in \citet{gilmoreUVcp}.  We calculate the
reprocessing of ionizing radiation by the intergalactic medium (IGM)
using the radiative transfer code \textsc{cuba}
\citep{haardt&madau96}, and use the observed ionization state of the
IGM to constrain our models.  The contribution of starlight to the EBL
is predicted by recent semi-analytic models (SAMs) of galaxy
formation, described in \citet{somerville08}.  These have been used by
\citet{gilmoreEBL} \citep[see also][]{primack08} to make predictions
for the evolving EBL.  Most previous modeling attempts, including our
recent work with this new SAM, have focused on observational data in
the optical and IR.  These are the wavelengths most relevant to
observations of relatively nearby ($z<0.5$) blazars with ground-based
instruments, which until recently have typically featured energy
thresholds above $\sim 150$ GeV.  With the recent launch of {\it
  Fermi} with its Large Area Telescope (LAT) sensitivity range of 20
MeV to 300 GeV, as well as the advent of new ground-based experiments
such as MAGIC-II with energy thresholds $<100$ GeV, it is now
important to make theoretical predictions of the UV background at
ionizing and non-ionizing wavelengths out to high redshift.  This
paper is an attempt to specifically target absorption in this region
of the gamma-ray spectrum.

Understanding the absorption that occurs for gamma rays observed
between 1 and 100 GeV is an uncertain undertaking due to the lack of
sensitive observations of the EBL at the corresponding UV wavelengths.
Moreover, the declining opacities for gamma rays in this region means
that sources are likely to be visible out to large redshift.
Evolution of the background must be taken into account when
calculating absorption for all but the nearest blazars, and at high
redshifts the EBL can have a spectral energy distribution (SED) much
different than observed locally.  The most distant object with
confirmed redshift that has been detected at VHE energies is currently
the flat-spectrum radio quasar 3C279 \citep{albert08} at
$z=0.536$. This object was observed at energies between 90 and 500
GeV, with a steep spectrum that was likely due in part to EBL
absorption. In the 10 to 100 GeV energy decade that is now being
probed by {\it Fermi} and upgraded ground experiments, the
characteristic redshift at which the EBL becomes optically thick to
pair-production is expected to increase to redshifts of several.

A small number of calculations have been performed that specifically addressed the
question of the gamma-ray absorption by the UV background.  In
\citet{madau&phinney96}, two different models of star-formation, based
on different assumptions about the B-band normalization, were used to
predict gamma-ray opacities from 10 to 200 GeV, with propagation of
ionizing photons through the IGM taken into account. This work
suggested that the universe becomes optically thick at a few tens of
GeV for gamma rays emitted at $z\sim 2$.  A second work which focused
on the UV background, \citet{oh01}, argued that the absorption by
ionizing photons was negligible, and that $<20$ GeV observed gamma
rays would only be significantly attenuated at higher redshifts, where
they would interact with photons below the Lyman limit. Ly$\alpha$
photons were found to be a significant component of the UV flux.  This
paper also explored the possibility of using {\it Fermi} to detect an
evolving blazar attenuation edge, which would probe high-redshift
star-formation.  Finally, the background model of \citet{salamon&stecker98} targeted absorption of 10--500 GeV gamma rays, and used an estimate of high-redshift star-formation based on evolution seen in damped Ly$\alpha$ systems.  This work also included a UV contribution from quasars.

At energies below the Lyman limit, lower bounds on galaxy emissivity
exist from number counts by {\it GALEX} \citep{xu05} and {\it Hubble
  Space Telescope} ({\it HST}), as well as balloon-based experiments
\citep*{gardner00}.  Such experiments are subject to systematic errors
in completeness and photometric measurement of apparent magnitude, and
can only test the background out to moderate redshift.  At higher
redshifts we no longer have measurements that directly connect to the
EBL, such as direct number counts and absolute photometry, and
uncertainties and possible biases in cosmological measurements such as
luminosity functions and star-formation rate density become increasingly
problematic.  

Measurements of the ionization state of the IGM can
provide constraints on ionizing flux. At redshifts higher than the
`breakthrough redshift' $\approx 1.6$, the universe is optically thick
to Lyman continuum photons, and ionizing fields become local, with a
mean free path that decreases rapidly at larger redshifts
\citep*{madau99}, while below this redshift the mean free path becomes
longer than the horizon length. Studies of the opacity of Ly$\alpha$
and other redshifted absorption lines place constraints on the
emission of UV photons by probing the neutral fraction, and therefore
the balance between photoionizations and recombinations
\citep{haehnelt01,madau04}.  As these lines are affected by the local
radiation, they provide information about sources existing at
approximately the redshift of the absorber \citep{haardt&madau96}.

Two methods of determining the ionization state of the IGM include the
proximity effect, in which one searches for the decrease in Ly$\alpha$
emissions near an AGN \citep*{dallaglio08,liske01}, and flux decrement
analysis, which utilizes hydrodynamic simulations to model the
distribution of Ly$\alpha$ absorption along the line of sight to an
AGN \citep[e.g.][]{bolton05}.  The line-of-sight proximity effect
utilizes the decrease in absorption lines in the vicinity of a quasar,
compared to farther away along the line of sight, due to increased
ionization fraction.  As the quasar has a known UV luminosity, the
deficit of absorption in this region can be used to estimate the
background; a larger change indicates a lower background flux.  As
quasars do not reside in typical cosmological environments, a number
of potential biases exist.  Quasars tend to be found in overdense
environments, which can lead to overestimates of the background flux
by as much as a factor of 3 \citep{loeb&eisenstein95}.  Time variation
in luminosity on the time-scale of photoionization, typically $\sim
10^4$ years, will also tend to bias results towards a high background,
as quasars tend to be selected in their brightest phases
\citep*{schirber04}.  It is also now recognized that using broad
emission lines such as Ly$\alpha$ tends to lead to underestimated
redshifts and therefore higher quasar luminosity \citep{richards02}.
This may have been a problem in many determinations of the proximity
effect.  The assumed cosmological model also affects the resulting background inferred by these measurements.  The second method mentioned, the less-direct flux decrement
technique \citep{rauch97}, is not without its own potential biases; it
relies on correct cosmological parameters and knowledge of the
quasar's unabsorbed continuum level, a problem at high redshift where
absorption is strong.  Newer attempts to correct for the biases in proximity effects measurements, such as \citet{dallaglio08}, have found lower values for the ionizing background flux that are more consistent with the flux decrement technique.

Observations of the Ly$\alpha$ forest can also provide clues about the
types of sources producing the ionizing background, which in our model
include star-forming galaxies and quasars.  The quasar luminosity
function (LF) has been measured by large-scale surveys such as
the Two-degree Field (2dF) \citep{boyle00} and the Sloan Digital Sky
Survey (SDSS) \citep{jiang06,richards06,croom04}, and data are also
available at a variety of frequencies from experiments such as {\it
  XMM}, {\it Chandra}, and {\it Spitzer}
\citep{barger05,matute06,brown06}.  The hydrogen of the intergalactic
medium (IGM) is known to be fully ionized below a redshift of $\sim$6
\citep{fan06}. Photons above the Lyman limit are responsible for
reionizing the universe and maintaining it in a highly ionized state.
The relative contributions of star-forming galaxies and AGN to this
process are not fully understood, but there is evidence that quasars
are a sub-dominant component at this epoch.  The decline of the quasar
luminosity function observed beyond redshift three constrains the quasar
contribution to the ionizing background to be $\sim 10^{-2}$ \citep{fan01,madau99}, unless there
is an unexpected steep upturn in the quasar luminosity function at low
luminosities.  A new approach by \citet{srbinovsky&wyithe07} utilizing
semi-analytic modeling sets limits on the quasar contribution to ionizing radiation of 1.4 to 14.5 per cent at $z=5.7$, and studies of the soft X-ray background also constrain this fraction to be subdominant \citep*{dijkstra04}.


Increased quasar emission is believed to be responsible for He II
reionization, which as tracked by He II Ly$\alpha$ absorption takes
place at a lower redshift than hydrogen, z $\sim$ 3
\citep[e.g.][]{bolton05}.  The shape of the ionizing background
therefore evolves in redshift, with a hardening of the spectrum that
is indicative of an increased contribution from quasars.  The degree
to which AGN dominate the UV background at the time of He reionization
is a debated issue, with some suggestions that stars and AGN provide
roughly equal contributions to the background at z $\sim$ 3
\citep{kriss01,smette02}.  One of the major sources of uncertainty in
this transition lies in the unresolvable faint end of the AGN
luminosity function \citep{schirber&bullock03}.  The ratio between
hydrogen and helium ionization fractions, particularly H I (13.6 eV)
and He II (54.4 eV), can be used to measure the slope of the total UV
spectrum in this regime.  The decrease in the optical depth of He II
indicates that the harder radiation from quasars increases with time
between z = 5 and 3 \citep{shull04,fan06}.

Aside from quasars, the known dominant sources of UV radiation are
short-lived massive stars, mainly of O- and B-type, which closely
trace the star formation rate density.  Estimating the ionizing
contribution from star-forming galaxies directly is complicated by the
fact that only a small fraction $f_{esc}$ of this radiation escapes
from galaxies due to neutral gas and dust in the interstellar medium,
as we will discuss in Section 2.4.


Because of the uncertain nature and evolution of sources of ionizing
radiation, in this paper we consider four models that attempt to span
a realistic range of assumptions.  In Section 2 we discuss the
inputs to our model, including a short description of our
semi-analytic models, assumed quasar luminosity density, and radiative
transfer code.  In Section 3, we introduce our four UV background
models and present results, including the evolving background
radiation and comparisons with Ly$\alpha$ forest measurements.  The
main results of the paper, gamma-ray opacities, are presented in
Section 4, with a discussion following in Section 5.  Our four models
are summarized in Table \ref{tab:models}, and their successes and
failures in accounting for the data are summarized in Table
\ref{tab:modelsum}.

\section{Modeling}

To calculate the evolving UV background we have used predictions of
the UV luminosity density from galaxies, as provided by recent
semi-analytic models (SAMs) of galaxy formation, combined with
estimates of quasar emissivity.  The combined emissivities from galaxies and quasars are integrated over redshift to find the evolving background flux.  Photons from these sources at energies above the Lyman limit can be absorbed and reradiated by the IGM; we calculate the effect of these processes using the \textsc{cuba} radiative transfer code.   In this section we discuss the
semi-analytic models and radiative transfer code used in calculating
the background, and show some general results.

\subsection{The semi-analytic models}

The semi-analytic models used in calculating the EBL in this work are
described in detail in \citet[][S08]{somerville08}, and are based on the
models described in \citet{somerville&primack99} and
\citet*{somerville01}, with several new updates and capabilities; we
provide a very brief summary in the following paragraphs.  Interested
readers should also see \citet{primack08} and \citet{gilmoreEBL},
which focused on the optical and IR EBL resulting from galaxies in
this model.

The backbone of the semi-analytic models are dark matter ``merger
trees'', which describe the cosmological assembly history of dark
matter haloes as they build up over time through mergers and accretion
of diffuse material. These merger trees may either be constructed via
Monte Carlo techniques based on the Extended Press-Schechter theory
(as in this work), or extracted from N-body simulations. We compute
the rate at which gas can cool via atomic processes and be accreted
onto the central galaxy within the potential well of each of these
collapsed and virialized DM haloes. Cold gas is then converted into
stars in both a ``quiescent'' and starburst modes. The ``quiescent''
mode takes place in isolated discs and is modeled according to an
empirical prescription based on the Schmidt-Kennicutt law
\citep{kennicutt98}. The starburst mode is triggered during mergers,
with gas being rapidly converted into stars on time-scales determined
from hydrodynamic simulations of galaxy mergers. Feedback from
supernovae and massive stars can reheat cold gas and expel it from the
galaxy. Chemical evolution is modeled using a simple instantaneous
recycling approximation with the effective yield treated as a free
parameter.

The most recent models also treat the growth of supermassive black
holes within galactic nuclei and the impact of the released energy on
galaxies and their environment. Seed black holes, perhaps originating
from the remants of very massive Pop III stars, are planted in the
top-level haloes in the merger trees, and can grow by two accretion
mechanisms. Galaxy mergers trigger rapid (Eddington limited) ``bright
mode'' accretion and correspond observationally to classical X-ray or
optically luminous quasars and AGN. Bright mode AGN activity drives
winds that can expel cold gas from the galaxy. Accretion of hot gas
from the hot halo feeds low-level (Bondi) ``radio mode'' accretion,
which is associated with the production of giant radio jets. These
jets are assumed to be capable of heating the surrounding hot gas
halo, offsetting or even quenching cooling flows. Although the radio
mode feedback in particular has been shown to play a key role in
reproducing important galaxy properties in these models, such as
galaxy colour bimodality and luminosity functions, the predictions for
observable properties of quasars and AGN in these models has not yet
been thoroughly tested. Therefore, in this work, we instead add in the
contribution to the background radiation due to quasars empirically.

The star formation and chemical enrichment histories for each galaxy
are convolved with stellar population models to produce synthetic
spectral energy distributions (SEDs). We make use of the \citet{bruzual&charlot03} models with a Chabrier stellar initial mass function.

Our models include a simple estimate of the attenuation of starlight
due to dust, based on a two-component model similar to that proposed
by \citet{charlot&fall00}. One component is the diffuse ``cirrus''
dust in the disc and another is associated with the dense `birth
clouds' surrounding young star forming regions. The $V$-band, face-on
extinction optical depth of the diffuse dust is given by
\begin{equation}
\tau_{V,0}\propto \frac {\tau_{\mathrm{dust,0}} \: Z_{\mathrm{cold}} \: m_{\mathrm{cold}}}{(r_{\mathrm{gas}})^2},
\end{equation}
where $\tau_{\mathrm{dust,0}}$ is a free parameter,
$Z_{\mathrm{cold}}$ is the metallicity of the cold gas,
$m_{\mathrm{cold}}$ is the mass of the cold gas in the disc, and
$r_{\mathrm{gas}}$ is the radius of the cold gas disc. To compute the
actual extinction we assign each galaxy a random inclination and use a
standard `slab' model. Additionally, stars younger than $10^7$ yr are
enshrouded in a cloud of dust with optical depth
$\tau_{\mathrm{BC,V}}=\mu_{\mathrm{BC}}\, \tau_{V,0}$, where
$\mu_{\mathrm{BC}}=3$. Finally, to extend the extinction correction to
other wavebands, we assume a Galactic attenuation curve \citep{cardelli89}
for the diffuse dust component and a power-law extinction curve
$A_{\lambda}\propto(\lambda/5500\mbox{\AA})^n$, with $n=0.7$, for the birth
clouds. The free parameters are adjusted to reproduce the observed
ratios of far-UV to far-IR luminosity in nearby galaxies.

We consider two different choices of parameters for our semi-analytic
models, which differ primarily in the assumed cosmological
parameters. The free parameters that control galaxy formation in each
model are then tuned to match local galaxy observations, as described
in S08 (the actual values of the parameters for both models are also
given in S08, Table 2).  The `fiducial' model adopts a concordance cosmology
with $\Omega_m$ = 0.3, $\Omega_{\Lambda}$ = 0.7, $h$ = 0.70, and
$\sigma_8$ = 0.90.  Our `low' model adopts the best fit values from
WMAP3 for these parameters, with $\Omega_m$ = 0.2383, $\Omega_{\Lambda}$ = 0.7617, $h$ = 0.732, and
$\sigma_8$ = 0.761. The most relevant difference in this work is the value of
the power spectrum normalization $\sigma_8$.  The lower
normalization of the primordial power spectrum in the `low' model leads to delayed
structure formation and  decreased luminosity
densities at high redshifts (see S08). Adopting the best-fitting
parameters from the more recent analysis of WMAP5, which favored
$\sigma_8=0.82$ \citep{komatsu09}, gives nearly identical results to
our WMAP3 (low) models. 

\begin{figure}
\psfig{file=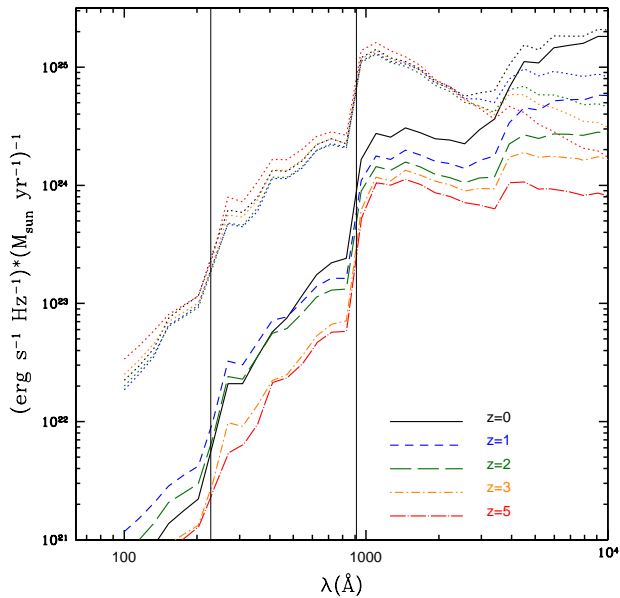,width=\columnwidth}
\caption{The emissivity due to galaxies predicted by our fiducial
  galaxy-formation model at a number of redshifts, normalized to 1
  $M_\odot$/yr. Dotted curves show the emission predicted in the
  absence of dust extinction.  Vertical lines indicate the ionization
  energies of H I and He II at 912 and 228 \AA.}
\label{fig:starsed}
\end{figure}

The predictions for attenuated and unattenuated emissivity from our
fiducial semi-analytic model are shown for several redshifts in
Fig.~\ref{fig:starsed}.  Predictions from the `low' model are
qualitatively similar, although because of the delayed star formation,
galaxies tend to have higher gas surface densities, and therefore
higher dust opacities and larger attenuation values are predicted.

\begin{figure}
\plottwo{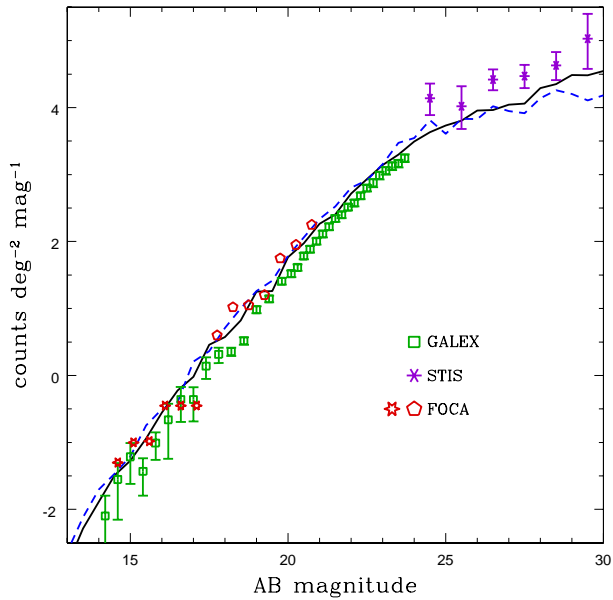}{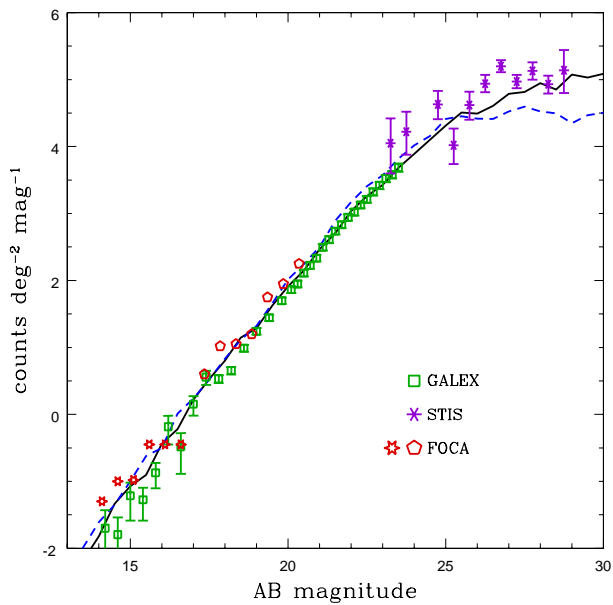}
\caption{Number counts in the {\it GALEX} 1530 \AA~(upper) and 2310 \AA~bands (lower).  The solid black line shows the fiducial model, and dashed blue shows the low model.  Note that the low model has counts equal to or slightly greater than the fiducial model at some magnitudes; this is due to differing amounts of dust extinction at low redshift between the two models.   Data are from {\it GALEX} \citep[][green squares]{xu05}, STIS on {\it  HST} \citep[][purple asterisks]{gardner00}, and the balloon-borne FOCA experiment \citep[][red stars and open pentagons respectively]{iglesias04,milliard92}.  Following \citet{xu05}, all counts have been converted to the {\it GALEX} bands by assuming a UV spectral slope of -0.8.}
\label{fig:countsUV}
\end{figure}

In our companion EBL paper (\citealp{gilmoreEBL}, see also \citealp{primack08}), we 
compare the predictions of our SAMs with a broad range of data,
including local luminosity functions, optical and IR luminosity
density, and number counts in a variety of bands from the UV to
far-IR.  In a planned future paper we will make a careful comparison
of these models with the galaxy population at high redshift.  Here we
show just a few representative results demonstrating that our models
are doing reasonably well at reproducing the UV properties of local
and distant galaxies. In Fig. \ref{fig:countsUV} we compare our model
predictions with galaxy number counts in two UV bands, using data from
the {\it GALEX} satellite and other experiments.  This provides a test
of the low-redshift normalization of our model in the UV range.  {\it
  GALEX} has surveyed the sky these bands and provided data
down to magnitude $\sim 23.5$ \citep{xu05}.  At fainter magnitudes,
there are measurements from the STIS instrument on the {\it HST}
\citep{gardner00}, albeit with large uncertainty due to poor
statistics.  Populations of brighter objects have also been probed by
the FOCA balloon-borne UV telescope, and counts from this instrument
at 2000 \AA~have typically yielded higher numbers than {\it GALEX}
after wavelength correction, possibly due to differences in calibration.  Our
models show good agreement with the data at 2310 \AA, but are a bit
higher than the {\it GALEX} observations at 1530 \AA, though they are
not in disagreement with the FOCA data.  

\begin{figure}
\psfig{file=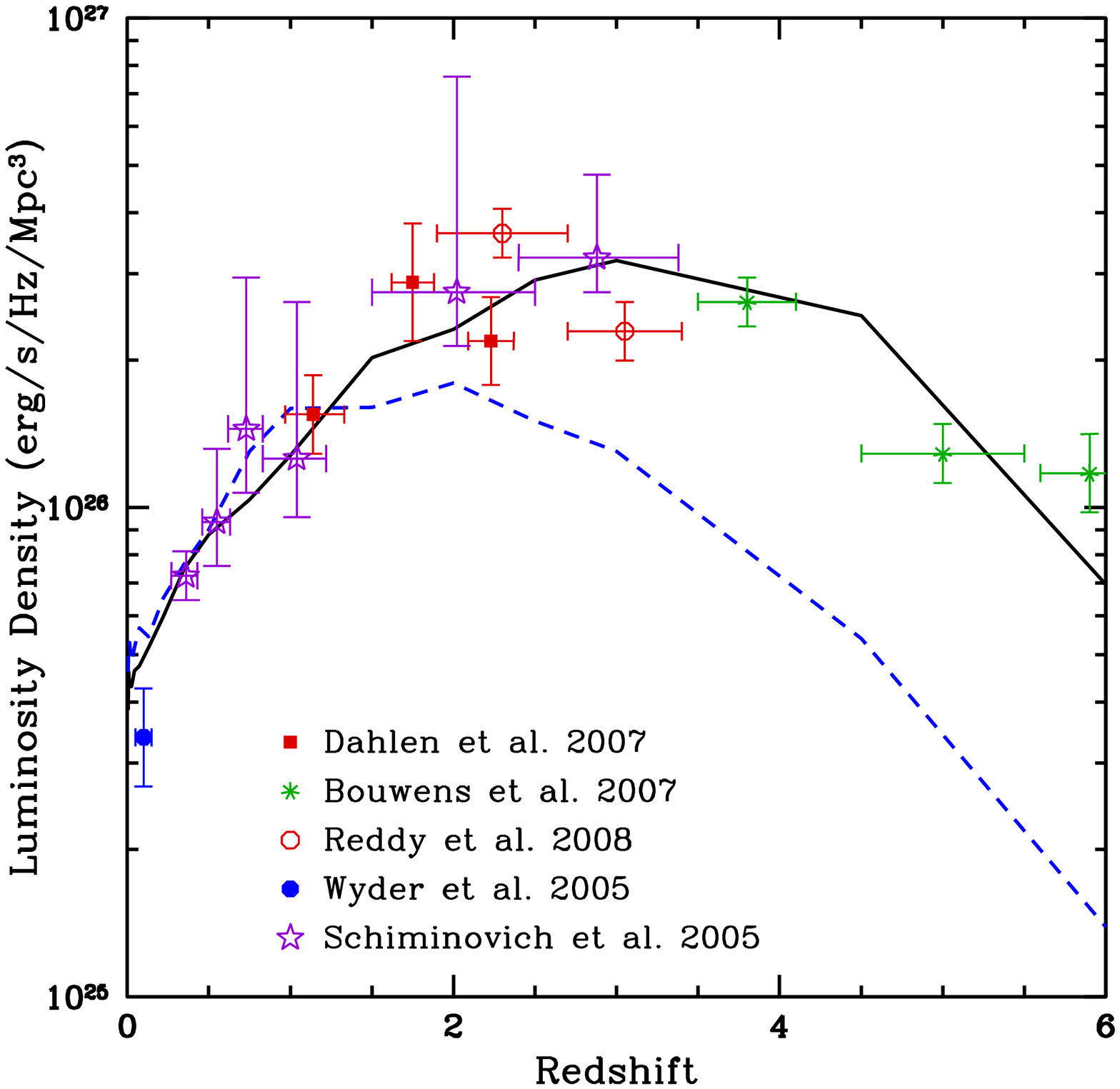,width=\columnwidth}
\caption{The emissivity at 1500 \AA~as a function of redshift in our models.  As previously, the solid black line is the fiducial model, and the dashed blue line shows the low model.  The blue circle at redshift 0.1 is GALEX data from \citet{wyder05} and the purple stars are measurements using GALEX and other data \citep{schminovich05}.  The red squares are from GOODS \citep{dahlen07}, the red circles are determinations from ground-based observations \citep{reddy08}, and the green stars are from \citet{bouwens07}. }
\label{fig:ldevo1500}
\end{figure}

Recent data from a variety of instruments has constrained the UV luminosity density out to high redshift.  In Fig. \ref{fig:ldevo1500} we have compared the UV emissivity in galaxies from our models against data at a rest frame emission wavelength of approximately 1500 \AA.  We find that data from the {\it GALEX}-VVDS, GOODS, and deep {\it HST} ACS imaging all agree reasonably well with the UV evolution of our fiducial model.

\subsection{Star formation}
\label{sec:sfr}

\begin{figure}
\psfig{file=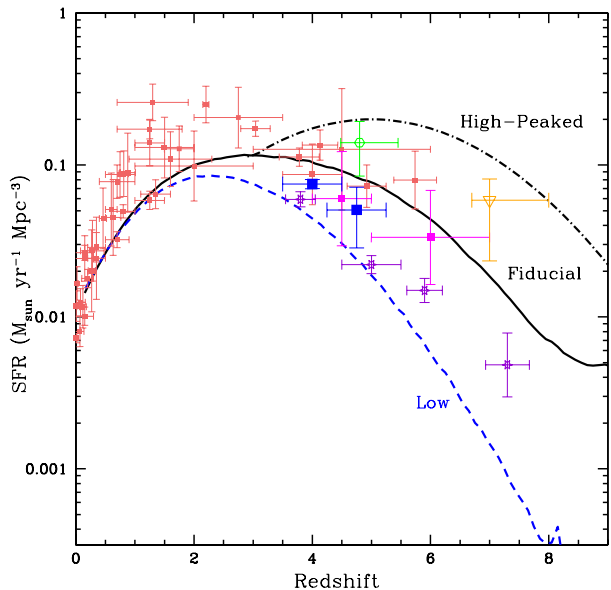,width=\columnwidth}
\caption{The global star formation rate density predicted by our
  models, compared with a compilation of observational data.  The
  solid black and dashed blue curves show the SFRD history of our
  fiducial and low models, respectively.  The black dash-dot
    curve which diverges from the fiducial curve above redshift three
    is the `high-peaked' form which we discuss in the text.  The red
  squares at lower redshift are from the compilation of
  \citet{hopkins04}.    The purple stars are from
  observations by \citet{bouwens08,bouwens07} of dropout-selected
  galaxies.  For these we show the dust-corrected results from integrating the luminosity functions down to a value of 0.04 L$_*$ at $z=3$; it is possible that fainter objects provide an additional contribution.  The magenta squares at redshift 4.5 and 6 show inferred
  star formation rates from gamma-ray burst observations
  \citep{yuksel08}.  The green circle is based on observations of
  Lyman-break galaxies at $z\sim 5$ \citep{verma07}, and the orange
  triangle is an upper limit from VLT data \citep{mannucci07}.   The blue squares are results from the Subaru Deep Field \citep{ouchi04}.  All data have been corrected for extinction (by the authors) and converted to a Chabrier IMF.}
\label{fig:sfr}
\end{figure}

The star-formation rate density (SFRD) as a function of redshift in
the `fiducial' and `low' models are shown in Fig.~ \ref{fig:sfr},
compared with observational estimates of star formation density at
various redshifts, all of which has been converted to a Chabrier
initial mass function (IMF).  At $z < 1$, both of our models are in good agreement with the
observational compilation of \citet{hopkins04}, while at $1<z<2$ they
tend to skirt the lower envelope of observational values. However,
there are still large discrepancies between SFR estimates from
different indicators and different data sets at these redshifts, in
part due to the increased fraction of star formation in
heavily obscured systems \citep[e.g.][]{hopkins07a}, where the
correction for dust obscuration is uncertain.  At $z>2$, the SFRD in
the `low' model declines fairly steeply, while in the fiducial model
the SFRD remains nearly constant from $2 < z < 5$ and then declines
more gradually. As discussed above and in S08, this is because of the
lower normalization of the power spectrum and reduced small scale
power in this model, which delays the formation of structure.

Above redshift four, observational estimates of global star-formation
rates diverge, and different measurements can disagree by as much as
an order of magnitude. Studies of UV luminosity functions of dropout
galaxies by Bouwens and collaborators \citep{bouwens08,bouwens07} find
relatively low values for the global SFR, with a monotonic decrease
above redshift four. Higher rates have been found by other authors,
including those who have derived star-formation history from
detections of gamma-ray bursts \citep{yuksel08}. These studies
suggest a much higher rate of star formation which does not decrease
significantly until $z > 6$. This may be due in part to the fact that
the Bouwens et al. data points that we report here were obtained by
integrating the UV luminosity function down to a luminosity
corresponding to 0.04 times the observed value of $L_*$ at redshift
three. Other authors make different choices for the lower limit of
integration, and the relatively steep slope of the UV LF at these
redshifts implies that this can make a significant difference. The SAM
predictions shown include the star formation in all galaxies (down to
the mass resolution of the simulation).

Our `low' semi-analytic model, based on the lower determination of
$\sigma_8$ in WMAP3, produces a star-formation rate that is lower than
most of the data points at mid and high redshift, and reproduces the
rapid fall-off in star formation indicated by the Bouwens points.  Our
`fiducial' model, based on WMAP1, does a better job of matching the
higher star-formation rates seen in other dropout analyses, as well as
the data from gamma-ray bursts.  

Although not shown here, another way to constrain the star formation
history of the Universe is via the build-up of stellar mass in the
form of long-lived stars. As shown in S08, the fiducial model predicts
a stellar mass density that is higher than observational estimates by
a factor of two at $z=1$ and a factor of three at $z=2$, while the
`low' (WMAP3) model produces good agreement with the stellar mass
assembly history. As discussed in S08, this discrepancy between the
observed stellar mass assembly history and the star formation history
has been pointed out in a number of recent papers
\citep[e.g.][]{dave08,fardal07}, and one possible explanation is that
the stellar IMF was more top-heavy in the past (so that more UV
photons were produced per total unit mass of star formation). However,
it is also still possible that the discrepancy is simply due to
observational uncertainties in the estimates of star formation rates
and/or stellar masses.

In addition to the star-formation histories predicted self-consistently in our fiducial and low semi-analytic models, we consider an additional ad hoc ‘high-peaked fiducial’ form for the SF history above $z=3$. This is not a semi-analytic model; it is simply a functional form that was chosen to be consistent with the highest observational determinations of the star-formation rate.  We then utilize the same redshift-dependent dust extinction factors as the fiducial model. We include this case to illustrate the predictions for gamma-ray attenuation for an extreme model with the maximum plausible UV background at high redshift. However, we note that as the fiducial model already produces an integrated stellar mass density in excess of that observed at high redshift, the high-peaked model is strongly disfavoured by these observations.

\subsection{Radiative transfer}

Ionizing photons from galaxies and quasars which escape into the intergalactic
medium (IGM) are processed by neutral hydrogen and neutral and
singly-ionized helium which resides in Ly$\alpha$ forest clouds (LAC)
and thicker Lyman-limit systems (LLS), defined here as having column
densities $>10^{17.2}$ $\mbox{cm}^{-2}$ .  This has a strong effect on
the spectrum and intensity of the average background field.  The
propagation of ionizing flux through the IGM in our models is
calculated using an updated version of the \textsc{cuba} code.  An
earlier version of \textsc{cuba} is described in
\citet{haardt&madau01} and is based on the theory of
\citet{haardt&madau96}.  Here we briefly summarize some of the main
ideas and formalism from these papers.

The effect of residual neutral gas on the ionizing radiation field can
be described in general terms by the radiative transfer equation:

\begin{equation}
(\frac{\partial}{\partial t}-\nu \frac{\dot{a}}{a}\frac{\partial}{\partial \nu}) J =-3\frac{\dot{a}}{a}J-c\kappa J+\frac{c}{4\pi}\epsilon.
\end{equation}

\begin{figure}
\psfig{file=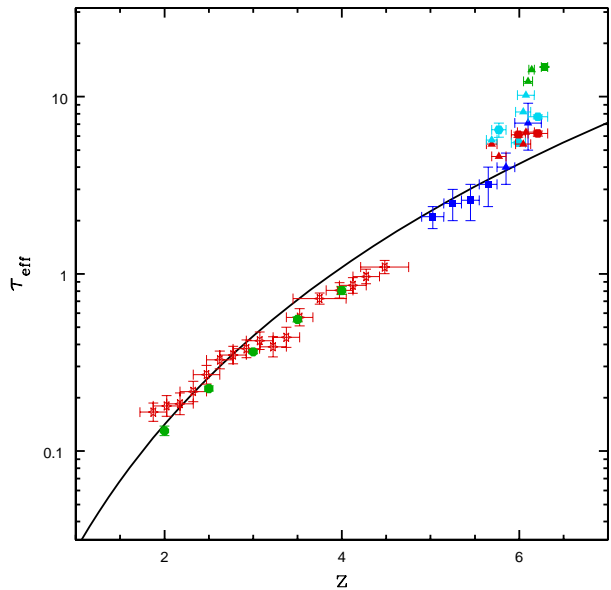,width=\columnwidth}
\caption{Effective optical depth as a function of redshift from our
  assumed absorption cloud distribution.  Data from quasar spectra are
  shown at ($5<z<6.5$) from \citet{fan06}; here the blue points are
  averaged Gunn-Peterson measurements, and the red, cyan, and green
  symbols are Ly$\alpha$, $\beta$, and $\gamma$ measurements of the
  highest-redshift individual objects.  Values at lower redshifts are
  from \citet{schaye03} (green circles) and \citet{dallaglio08} (red
  stars). }
\label{fig:tauhi}
\end{figure}

\noindent Here $J(\nu)$ is the intensity of the radiation field for
frequency $\nu$, $\epsilon(\nu)$ is the emissivity, $a$ is the
cosmological scale factor, $c$ is the speed of light, and $\kappa$ is
the continuum absorption coefficient.  This equation accounts for both
the redshifting of photons to lower energies, and absorption by
neutral gas.  Quasars and star-forming galaxies contribute to
$\epsilon(\nu)$ in our model, along with the diffuse reemission of
absorption systems.  Lyman absorbers are taken to have a distribution
that can be described in terms of power laws in column density and
redshift
\begin{equation}
\frac{\partial^2 N}{\partial N_{HI} \partial z} \propto N_{HI}^{-1.5} (1+z)^\gamma
\label{eq:lyforest}
\end{equation}
with parameters
\[\gamma = 0.16 \mbox{ (LAC, } 0<z<1.4) \]
\vspace{-0.4cm}
\[\gamma = 3.0 \mbox{ (LAC, } 1.4<z) \]
\vspace{-0.4cm}
\[\gamma = 1.5 \mbox{ (LLS, all redshifts) } \]

\noindent used in these models.  A distribution with slope -1.5 in column density has been shown to describe absorbers over a wide range in $N_{HI}$ \citep{hu95}, and the slopes for redshift evolution are based on observational determinations \citep{kim97,stengler95,bechtold94}.  The effective optical depth from the
Ly$\alpha$ forest absorption in this distribution is shown in Fig.~
\ref{fig:tauhi}.  Note that our Ly$\alpha$ optical depth does not
follow the upturn seen at $z\sim 6$, where a rapid rise in absorption
may signal that our assumption of a uniform UV background is no longer
valid.

It should be noted that the exact form of the column distribution
function is not critical, as it is the integrated value of this
parameter from which we derive the effective optical depth and
therefore the average background.  The effective depth is dominated by
clouds with opacity near unity.  Using a power-law form simplifies the
integration process, and speeds up our computation with little loss in
accuracy.  We do caution readers that the choice of distribution
function can have a large effect on He II absorption, and in turn on
the background above 54.4 eV which we report in Fig.~
\ref{fig:eblhistUV}.  The background at these energies is not expected
to affect our gamma-ray attenuation significantly because the photon
density at these high energies is so low.

Lyman systems reradiate a fraction of the absorbed light via recombination radiation.  Our code accounts for the contribution of H I recombinations to the UV flux via free-bound, Ly$\alpha$, and two-photon continuum emission.  For the latter two, only the non-ionizing background is affected.  The total proper volume emissivity from IGM clouds from radiation released for a particular mode can be quantified as

\begin{equation}
\epsilon (\nu,z) = h \nu f_i(\nu) \: W_{abs}(z) \: \Xi (z,\nu) \, \frac{\alpha_i}{\alpha_{tot}} \frac{dz}{dl}
\end{equation}

\noindent where $\alpha_i$ is the fraction of recombinations leading
to the particular mode, which has probability $f_i(\nu)$ of creating a
photon of energy $\nu$.  In the case of Lyman-$\alpha$ emission this is
simply a delta function at the line energy, and for the continuum
distributions descriptions can be found in \citet{osterbrock89}.  The
remaining functions contain the details of emission and absorption
from absorption systems

\begin{equation}
 W_{abs}(z) = \int_{\nu_{th}}^\infty \frac{4\pi J(\nu',z)}{h \nu'} w_{abs}(\nu') d\nu' 
\end{equation}
\begin{equation}
\Xi (z,\nu) = \int_0^\infty \frac{\partial^2 N}{\partial N_{HI} \partial z} p_{em}(\nu,N_{HI}) d N_{HI}.
\label{eq:xi}
\end{equation}
The first of these quantities is the rate of ionizations by the background field $J(\nu',z)$.  Here $w_{abs}(\nu')$ encodes the information about the optical depth for photons of a given energy, and thus the probability of being absorbed.  Equation \ref{eq:xi} is the integral over absorption systems, multiplied by $p_{em}(\nu,N_{HI})$, which is the probability of a photon of given energy escaping from a cloud after emission.  The code does not include the contribution from sawtooth modulation due to H I and He II Lyman resonances \citep{madau&haardt09}.

We do not include radiation from recombinations to He I, as neutral
helium is small in number density compared to both H I and He II
(though that may not be the case at redshifts on the verge of
reionization).  As shown in \citet{haardt&madau96}, thermal
collisional effects can provide a sizable fraction (20 to 30 per cent)
of the H I emission by Ly$\alpha$ and two-photon processes; these are
not accounted for in our code.  Collisions between He II atoms are
never significant, as there is insufficient thermal energy to excite
these modes.

\subsection{Ionizing escape fraction from galaxies}

The fraction of ionizing photons produced that escape from galaxies
into the IGM is a free parameter in our model.  This parameter is poorly constrained, with observations and
simulations giving widely different and sometimes conflicting results.  In the literature, the escape fraction may be defined in a couple ways.  The absolute escape fraction is simply the fraction of radiation at wavelengths just shortward of 912 \AA~ which escapes the dust and neutral hydrogen in a galaxy.  This definition is most relevant for the purposes of modeling the ionizing background.  What is actually measured in observations is the relative escape fraction, where the ionizing flux is compared to a non-ionizing wavelength, often 1500 \AA.   As described below, we use a relative definition which separates the amounts of attenuation due to dust and H I (Equation \ref{eq:fesc}).  Ionizing radiation from quasars is not attenuated in escaping the host galaxy in our model.

Direct detection of escaping UV radiation has only been successful in
a handful of individual cases.   As two of these detections have been
for galaxies at z$\sim$3 with large escape fractions measured
\citep{shapley06}, the fact that many low redshift attempts to find
ionizing radiation have failed with low upper bounds may suggest
evolution in this quantity between redshifts one and three.  Rather
firm upper limits on escape fraction from direct detection efforts
exist for lower redshift galaxies (see the compilation in
\citealt{siana07}).  \citet{steidel01} reported ionizing flux from 29 stacked galaxies at $z\sim 3.4$, at a level indicating little or no attenuation.  However, this result suffered from a selection bias, as the Lyman--break galaxies used were chosen from the bluest quartile of the population.

While observations have mainly determined upper limits on the ionizing escape fractions, some authors have used the ionization state of the IGM to derive lower limits.  Ionization rates inferred from Ly$\alpha$ forest
data and reasonable extrapolations of source number to faint
luminosity can require a high escape fraction.  Values of $\gtrsim 20$
per cent above redshift 5 were found to be needed in
\citet{bolton&haehnelt07}.  \citet{srbinovsky&wyithe08} have found that constraints on the escape fraction from
$5.5 <$ z $< 6.0$ from N-body simulations require a global minimum
of 5 per cent to match Ly$\alpha$ data, with a higher
fraction needed in the event that star formation in galaxies in
smaller haloes is suppressed .

Recently, detailed Adaptive Mesh Refinement (AMR) N-body
hydrodynamical simulations of high-redshift galaxies ($3<$ z $< 9$) by
\citet*{gnedin08} have found low escape fractions of 1--3 per cent,
without strong evolution in redshift or dependence on galaxy
properties.  This work found that most escaping ionizing radiation
originated from stars in a thin shell at the outside of the H I disc.  Smaller galaxies have less escaping radiation due to the fact that their H I discs are thicker relative to the distribution of young stars.  Dust is found to have little effect on the
escape of Lyman-continuum radiation, as the unobscured minority of
stars that provide most of the escaping ionizing radiation have
essentially no attenuation due to dust, while stars that have
translucent ($\tau \sim 1$) optical depths due to dust are generally
completely obscured by HI.  Another analysis has been undertaken
using a smoothed particle hydrodynamics (SPH) code by
\citet{razoumov&sommerlarsen07,razoumov&sommerlarsen06} and has found
evolving escape fractions, with $f_{esc}= 6$ to 10 per cent
at $z=3.6$ decreasing to 1 to 2 per cent at $z=2.4$.  This simulation
did not include the effects of dust.

Our semi-analytic models predict the emissivity from star-forming
galaxies down to a minimum rest-frame wavelength of 100 \AA.
Lyman-continuum photons are attenuated by a factor $\feschi$; this
determines the absorption of photons shortward of 912 \AA~and is a
non-evolving input to our radiative transfer code.  This parameter is
defined as the following ratio of intrinsic and observable
luminosities at 912 and 1500 \AA
\begin{equation}
\label{eq:fesc}
\feschi = \frac{(L_{912}/L_{1500})_{\rm
    escaping}}{(L_{912}/L_{1500})_{\rm intrinsic}}=f_{\rm esc}
f^{-1}_{1500}
\end{equation}

\noindent where $f_{\rm esc}$ is the absolute attenuation factor from both dust
and H I for ionizing photons near the Lyman limit, and $f_{1500}$ is
the factor from dust alone at 1500 \AA.  As dust absorption is an
evolving effect included in the semi-analytic model, total absorption
$f_{\rm esc}$ for ionizing photons escaping from galaxies must be
interpreted as the product of \feschi\ and the average $f_{1500}$ for
a particular redshift in the model.  The average value of $f_{1500}$
in our model is higher at high redshift, and is higher in the low
model than in the fiducial model at high redshift.  Typical values of
$f_{1500}^{-1}$, as seen in the difference between attenuated and
unattenuated curves in Fig.~ \ref{fig:starsed}, vary from about 6.8 at
$z=2$ to 11.2 at $z=5$ for our fiducial model, and 9.7 to 14.2 in the
low model.  As noted in \citet{gnedin08}, taking the total absorption
to be a product of the dust and H I factors may not be physically
realistic, as the two components are may not be distributed in the
same way within the galaxy, however it is sufficient for the purpose
of calculating emissivities, as we are dealing only with global
quantities.

\subsection{Quasar emissivity} 

Quasar input to our model is accomplished using an assumed UV
luminosity density, which determines
the output at all energies via a fixed spectral form.  We have used
the quasar luminosity functions at 912 \AA~ determined by
\citet*[][HRH07]{hopkins07} which are based on a large observational
data set and spectral and obscurational modeling.  This work
found that, with the appropriate corrections for obscuration, a single
bolometric function could match data in each band.  Both the bright-
and faint-end slopes of the LF are argued to become more shallow at
higher redshift, indicating a increasing contribution from bright,
unobscured AGN, which dominate the total AGN luminosity above redshift
$\sim 2$.

Another approach to modeling the quasar contribution to the background
was presented in \citet[][SB03]{schirber&bullock03}.  This work
estimated the evolution of the unobserved faint-end of the quasar
luminosity function at high redshift using observational constraints
on the ionizing background. They found very different results
depending on whether they used constraints on the ionizing background
from the quasar proximity effect or from the Lyman-$\alpha$ flux
decrement method.  Assuming the broken power law universal quasar spectrum we present below, the HRH07 luminosity density evolution at 912 \AA~is
similar to Model `A' of SB03, which produced the lowest fluxes and was
found to be consistent with the flux decrement data when combined with
a substantial contribution from star-forming galaxies.  The highest
derived flux, arising in Model `C', was sufficient to produce ionizing
photons at the level suggested by proximity effect measurements.
While we recognize that it is probable that many of these proximity
effect measurements overestimate the background due to aforementioned
biases, we will use the SB03 Model `C' for the purposes of creating
and analyzing an extremely quasar-dominated background model.

\begin{figure}
\psfig{file=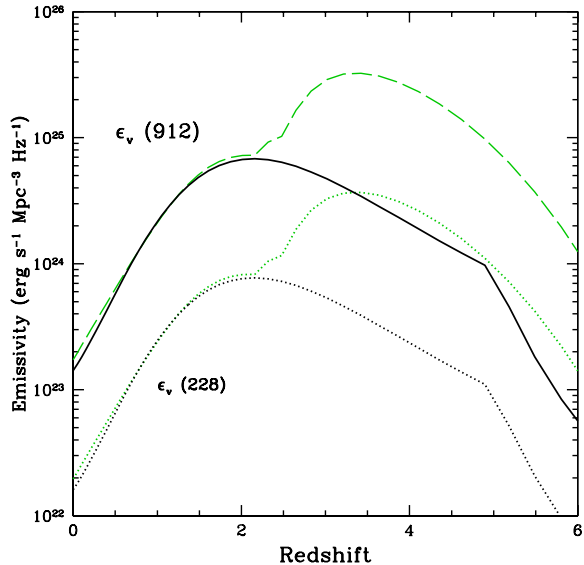,width=8.0cm}
\caption{Quasar luminosity density for the \citet{hopkins07} (solid
  black) and \citet{schirber&bullock03} Model `C' (broken green)
  models at 912 (upper lines) and 228 \AA~ (lower lines).  The latter
  has been multiplied by a factor of 0.8 to better match the
  observations at low redshift. }
\label{fig:emiss}
\end{figure}

Luminosity densities at the H I and He II Lyman limits for each of
these models are shown in Fig.~ \ref{fig:emiss}.  We have renormalized
the original model proposed by SB03 by a factor of 0.8 to better match
the HRH07 results below redshift 2.3.

Quasars are assumed to have a spectrum which can be modeled as a
broken power law in $F_\nu = dF/d\nu$.  In the extreme-UV, we have
adopted the hard spectrum suggested by \citet{telfer02}, from
observations with the Faint Object Spectrograph on {\it HST} of
quasars at a wide variety of non-local redshifts ($z>0.33$).  The
values we assume are as follows
\begin{equation}
F_\nu \propto \nu^{-\alpha}
\end{equation}
with indices


\[\alpha=0.4 \hspace{1cm} (12 \: \mu\mbox{m} < \lambda)\]
\vspace{-0.4cm}
\[\alpha=1.3 \hspace{1cm} (1 \: \mu\mbox{m} < \lambda < 12 \: \mu \mbox{m})\]
\vspace{-0.4cm}
\[\alpha=0.2 \hspace{1cm} (500 \: \mbox{nm} < \lambda < 1 \: \mu \mbox{m})\]
\vspace{-0.4cm}
\[\alpha=0.5 \hspace{1cm} (120 \: \mbox{nm} < \lambda < 500 \: \mbox{nm})\]
\vspace{-0.4cm}
\[\alpha=1.57 \hspace{1cm} (\lambda < 120 \: \mbox{nm})\]
\vspace{-0.4cm}
\[\alpha=0.9 \hspace{1cm} \mbox{(Soft X-rays,} >\mbox{500 eV)}\]

\noindent The contribution of quasars to the background is only
non-negligible in the extreme-UV ($\lambda <120$nm).  At other wavelengths we use slopes to roughly fit typical x-ray spectra of Seyfert 1 galaxies, and the optical and IR emission of quasars seen in SDSS and other observations \citep{vandenberk01,sanders89}.

\section{Cosmological Models and Resulting UV Background}

To calculate the evolving background radiation field, we have combined
three estimates of the star-formation rate density with two possible
forms for the quasar luminosity density.  The two quasar models are
the `realistic' estimate of \citet{hopkins07} from a large body of
observational data, and a higher `extreme' model motivated by
proximity effect measurements for $3<z<5$ from SB03.  Our four models
are summarized in Table \ref{tab:models}.

\begin{table}
\caption{The background models considered in this work.  The second
  and third columns show the star-formation histories and quasar
  luminosity densities used as inputs in each model.  The escape
  fraction in the last column refers to the values used in calculating
  the background flux and optical depth to gamma rays in Figs.~
  \ref{fig:eblhistUV} and \ref{fig:opdep_multi}, respectively.  Our star
  formation history scenarios are discussed at the beginning of this
  section.  `HRH07' refers to the best-fitting model of \citet{hopkins07},
  and `SB03 Model C' to the model in \citet{schirber&bullock03}.  We
  have multiplied the latter by a factor of 0.8 to better match the observed quasar luminosity density at low redshift. The escape fraction refers to the attenuation of ionizing photons from
star-forming galaxies by neutral hydrogen; attenuation by dust is
included intrinsically in our semi-analytic model.}
 \label{tab:models}
 \begin{tabular}{|lllc|}
\hline
{\bf Model} & {\bf SFR Density} & {\bf Quasar Luminosity} & {\bf \feschi }\\
\hline
\hline
1 & Fiducial & HRH07 & 0.1 \\
\hline
2 & Low & HRH07 & 0.2  \\
\hline 
3 & Fid. High-peaked & HRH07 & 0.1 \\
\hline
4 & Fiducial & SB03 Model C & 0.02 \\
\hline
\end{tabular}
 \medskip
\end{table}

\subsection{Ly$\alpha$ forest constraints}

\begin{figure}
\psfig{file=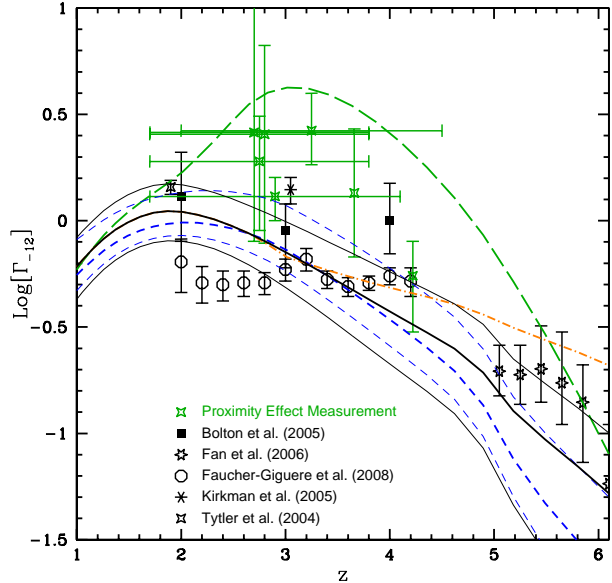,width=\columnwidth}
\caption{Ionization rate per hydrogen atom (with units of $10^{-12}$
  s$^{-1}$) in our four scenarios compared with data at a range of
  redshifts.  Black solid lines: fiducial model with H I escape
  fractions from star-forming galaxies of 0.02, 0.1, and 0.2 (bottom
  to top).  Dashed blue lines: low star-formation model, with escape
  fractions 0.1, 0.2, and 0.5.  Orange dash-dotted line: high-peaked
  star formation rate with escape fraction 0.1.  These aforementioned
  models all use the quasar emissivity of HRH07.  The long-dashed
  green line shows the fiducial SFR model with quasar model C of
  \citet{schirber&bullock03}, and escape fraction 0.02.  Data points
  are divided into those obtained from flux-decrement analysis (black)
  and those obtained via proximity effect near quasars (green).
  References for the former are
  \citet{bolton05,fan06,faucher08,kirkman05,tytler04}, and the latter
  include \citet{scott00}; \citet*{cooke97}; \citet*{giallongo97};
  \citet{cristiani95,williger94}; \citet*{lu91}; \citet*{bajtlik88}.
  Some points have been shifted slightly for readability.}
\label{fig:ionrates}
\end{figure}

Inferred ionization rates and column density measurements from the
Ly$\alpha$ forests of quasar spectra provide us with an independent
measurement of the UV background through its ionizing properties,
which we can compare with the IGM state computed by our radiative
transfer code.  In Fig.~ \ref{fig:ionrates}, we compare the ionization
rate (in terms of $\Gamma_{-12}$, the average rate per hydrogen ion
with units of $10^{-12}$ s$^{-1}$) with data from both the quasar
proximity effect and the flux decrement in Ly$\alpha$ forest
measurements.  As discussed in the Introduction, these two techniques
have tended to give disparate values for $\Gamma_{-12}$.  For the
fiducial and low models with the HRH07 QSO LF, we show ionization
rates with several values for the escape fraction of ionizing
radiation from the galactic H I disc.  With a moderate H I escape
fraction of 0.1 to 0.2, our fiducial model is able to reproduce the
level of ionizing background detected by most determinations using
flux decrement techniques.  Including attenuation by dust, this
corresponds to a total ionizing escape fraction of $\sim$ 1 to 3 per
cent, consistent with upper limits from observations as well as values
suggested by simulations \citep{gnedin08,razoumov&sommerlarsen07}.
With the low model, a higher H I escape fraction of $\sim 0.5$ is
necessary to match the highest redshift points ($z>5$), due to the
rapidly declining star-formation rate at high redshift. This escape
fraction is higher than suggested by some authors, but there are no direct constraints on escape fractions at such high
redshifts.

Based on the quality of fits for our different models to these flux decrement data, we have chosen escape fractions of 0.1 and 0.2
for our fiducial and low models, respectively, as reasonable values to
use in calculating the background and pair-production opacity.  Both of these models predict ionization rates which decline above
redshift 2.5; this is due both to the shape of the star-formation
history and the increasing opacity of the IGM with redshift.  

The flux decrement calculations from the largest quasar samples, \citet{faucher08} and \citet{bolton05}, find ionization rates that are essentially flat from $z=2$ out to $z=4$, albeit with differing normalizations.  Our model predictions are reasonably consistent with these observational estimates, considering the uncertainties involved. The high-peaked model better reproduces the flatness of the ionization rate from $z \sim$ 3 to 4, but still predicts too steep a rise with time from $z\sim$ 2 to 3. The SF history predicted by our fiducial model could be made perfectly consistent with the Faucher-Gigu`ere et al. (2008) data by assuming an escape fraction that evolves from $\sim 0.2$ at $z\sim 4$ to $0.02$ at $z\sim 2$.

The final scenario we examine uses the higher quasar emissivities of
Model `C' in \citet{schirber&bullock03}.  As the ionizing contribution
from star-formation is subdominant at all intermediate redshifts in
this case, we have assumed a low escape fraction of 0.02.  The
ionizing flux in this model is capable of reproducing the highest
measurements from the proximity effect.  We have already mentioned several known biases which may have artificially elevated these
values, and this model should be considered an extreme
possibility.

\begin{figure}
\psfig{file=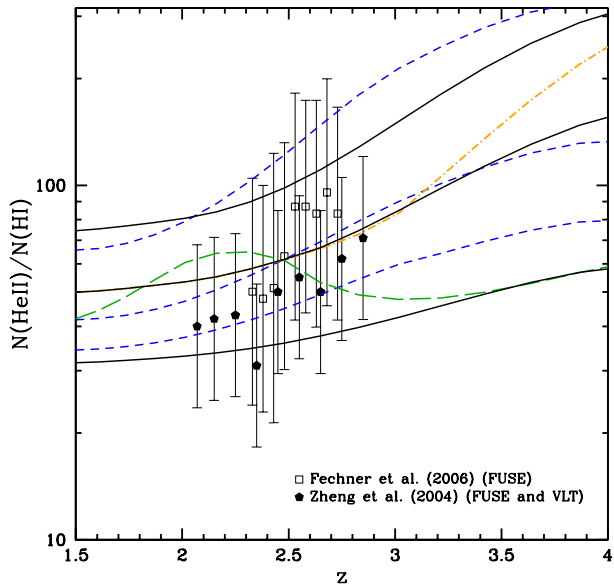,width=\columnwidth}
\caption{The ratio of He~II to H~I column densities plotted against
  redshift.  Higher values indicate a softer ionizing background, with
  comparatively more ionizing photons available per hydrogen atom.
  Line types in this plot are the same as in Fig.~\ref{fig:ionrates}.
  Data are from observations of He~II Ly$\alpha$ systems by
  \citet{zheng04} and \citet{fechner06}.  }
\label{fig:heii_hi}
\end{figure}

Another Ly$\alpha$ forest measurement which can provide insight into
the UV background is the relative abundance of H~I and He~II present
in the IGM.  This may be presented in terms of relative column
densities $N({\rm HeII})/N({\rm HI})$, or analogously as inverse
ionization rates for these components $\Gamma_{\rm HI} / \Gamma_{\rm
  HeII}$; this is often referred to as the UV softness parameter.  In
Fig.~ \ref{fig:heii_hi} we show how softness evolves with redshift for
each of our background models.  Our low, fiducial, and high-peaked
star-formation densities with the HRH07 QSO LF are able to provide a
reasonable match to observations when a moderate escape fraction is
assumed.  High escape fractions $\geq 0.5$, which are required for the
`low' SFR model to match ionization rates at high redshift, tend to
overpredict softness.  Our quasar-dominated model (SB03 model C) does
not reproduce the trend of increasing softness in the background field
with redshift, another factor which disfavors such a dominant
contribution from faint quasars.  The column density ratio in this
case is not found to be sensitive to the H~I escape fraction.

\subsection{The background flux}

\begin{figure}
\psfig{file=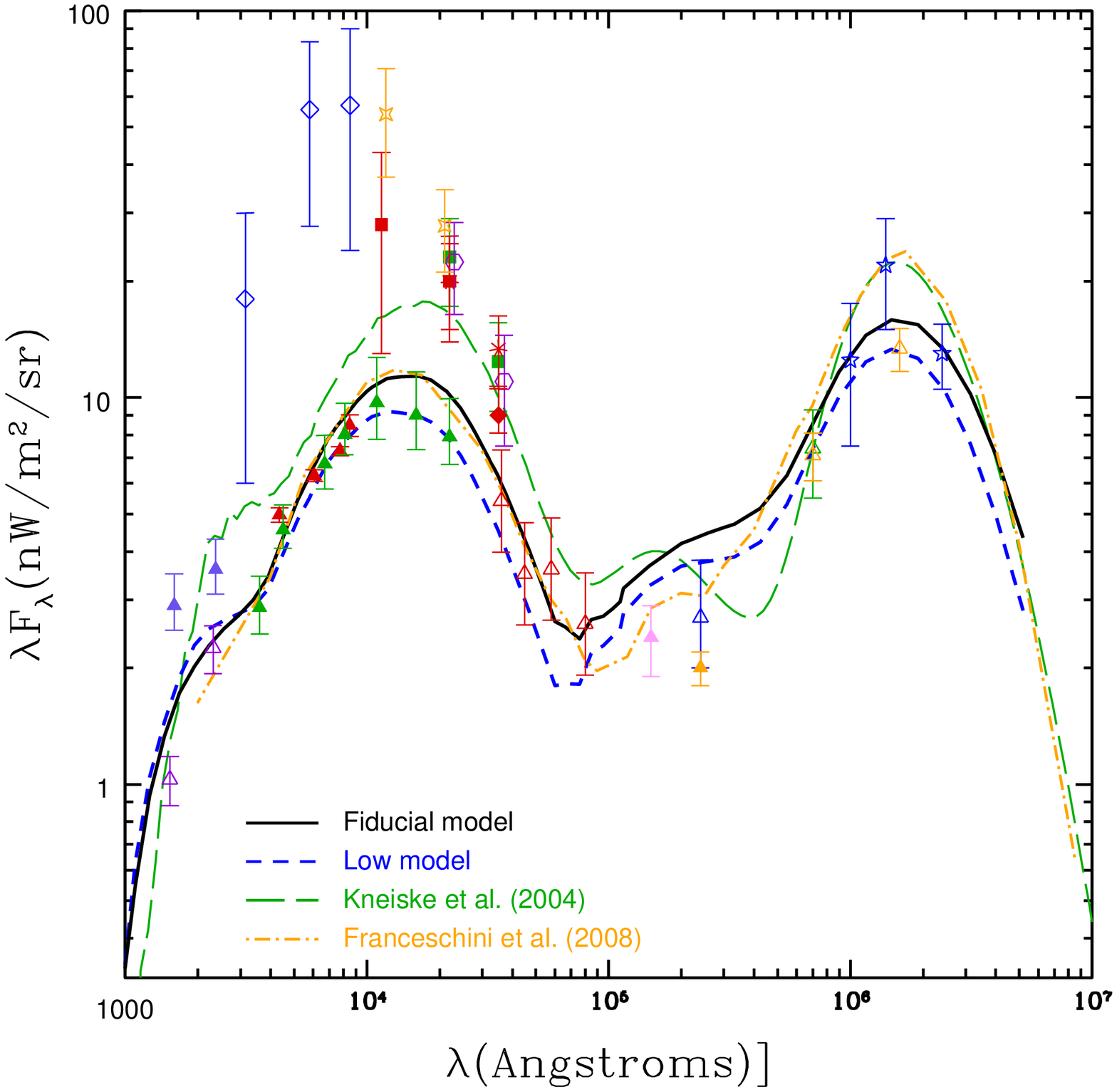,width=\columnwidth}
\caption{The predicted z=0 EBL spectrum from our semi-analytic models of galaxy formation from the UV to far-IR (see also \citealp{primack08}).  We show results for the fiducial (black) and low (dashed blue) models, compared with experiments at a number of wavelengths.  Other lines show the EBL models of \citet{franceschini08} and \citet{kneiske04} for comparison.  The blue-violet triangles are results from STIS on {\it HST} \citep{gardner00}, while the open magenta triangles are from {\it GALEX} \citep{xu05}.  The green and red triangles are from the Hubble Deep Field \citep{madau00} and Ultra Deep Field (Dolch, in preparation) respectively; the former also includes near-IR ground based-data.  Open red triangles are from IRAC on {\it Spitzer} \citep{fazio04}.  The blue diamonds are \citet{bernstein07} and other symbols in the near-IR are from several analyses of DIRBE data at 1.25, 2.2, and 3.5 $\mu$m (\citealp{levenson08,levenson07,cambresy01}, \citealp*{gorjian00,wright&reese00}).  In the mid-IR, counts data is shown from {\it ISO} \citep{elbaz02} and {\it Spitzer} \citep{papovich04,chary04,frayer06,dole06}.   In the far-IR, direct detection data is shown from DIRBE (\citealp{wright04}, blue stars).}
\label{fig:eblfluxUV}
\end{figure}

\begin{figure*}
\psfig{file=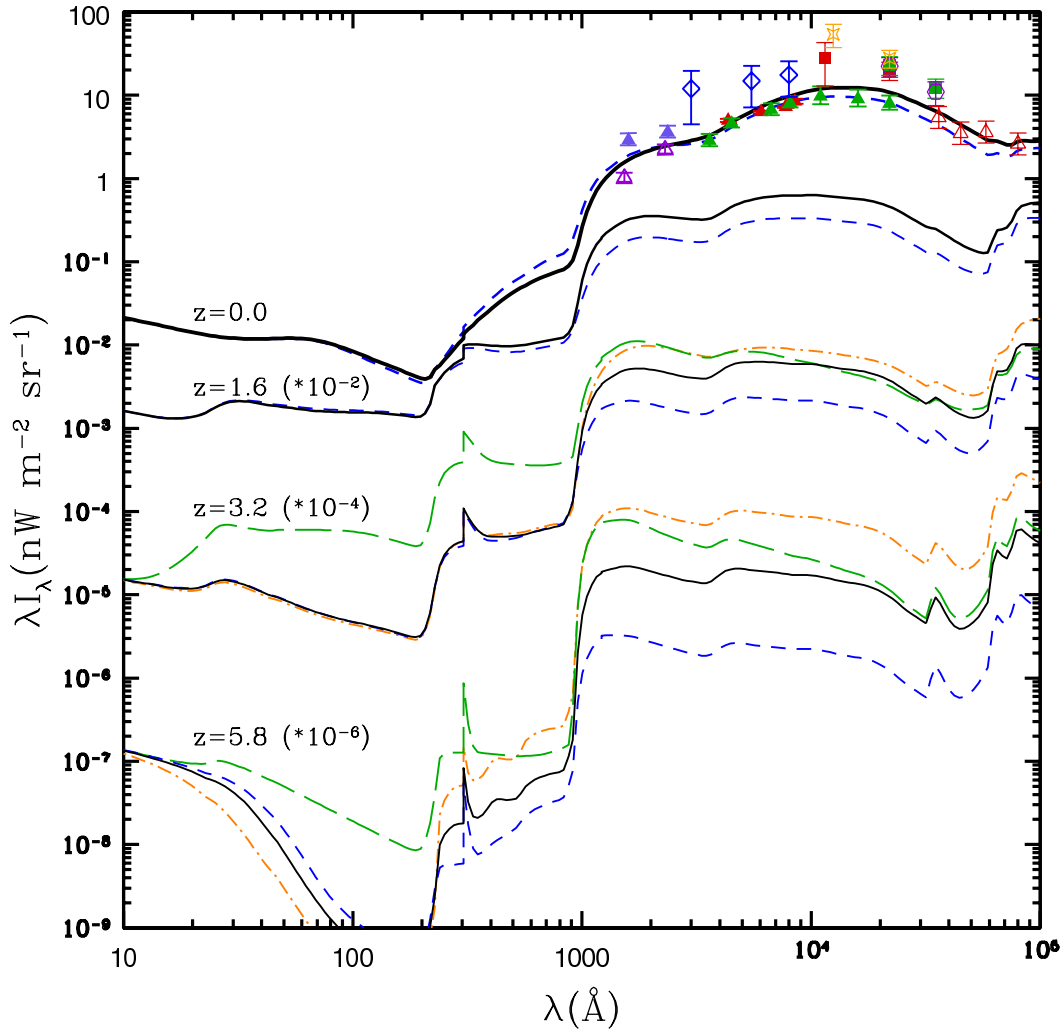,width=14cm}
\caption{The history of the background flux, shown at the present day
  and three other redshifts.  Intensities at the nonzero redshifts
  have been multiplied by the indicated factors ($10^{-2}$, $10^{-4}$,
  and $10^{-6}$ at $z=1.6$, 3.2, 5.8, respectively) to separate the
  lines.  Black solid line: fiducial model with H I escape fraction of
  0.1.  Dashed blue line: low star-formation model, with escape
  fraction 0.2.  Orange dash-dotted line: high-peaked star formation
  rate with escape fraction 0.1.  Green long-dashed: fiducial model
  with SB03 quasar contribution and escape fraction 0.02.  At low
  redshift, only the first two models are shown, as the other models
  do not produce discernibly different levels of background at these
  times.  We have also shown observational measurements of the
  background flux at $z=0$ in the UV, optical, and near-IR from Fig. \ref{fig:eblfluxUV}.  }
\label{fig:eblhistUV}
\end{figure*}

The key result of this work is a prediction of the evolving UV
background out to redshift $z\sim9$, which has been calculated from
our models for the total (stellar + quasar) emissivity $\epsilon(\nu,z)$ combined with a
calculation of the absorption and re-emission by IGM radiative
transfer processes.  Photons at non-ionizing wavelengths evolve passively, and are not attenuated significantly by any source.  For these photons, the flux at a redshift $z_0$ and frequency $\nu_0$ in proper coordinates can be written as \citep{peebles93}
\begin{equation}
J(\nu_0,z_0)=\frac{1}{4\pi} \int^{\infty}_{z_0} \frac{dl}{dz} \frac{(1+z_0)^3}{(1+z)^3}\epsilon (\nu,z) e^{-\tau_{eff}} dz,
\end{equation}
where $\nu=\nu_0(1+z)/(1+z_0)$ and $dl/dz$ is the cosmological line element, defined as
\begin{equation}
\frac{dl}{dz}=\frac{c}{(1+z)H_0} \frac{1}{\sqrt{\Omega_m(1+z)^3+\Omega_\Lambda}}
\label{eq:cosline}
\end{equation}
for a flat $\Lambda$CDM universe.  Here $\tau_{eff}$ is the optical depth for ionizing radiation due to Poisson-distributed HI absorption systems, defined as \citep{madau99}
\begin{equation}
\tau_{eff}(\nu_0,z_0,z)=\int_{z_0}^z dz' \int_{0}^{\infty} dN_{HI} \frac{\partial^2 N}{\partial N_{HI} \partial z'}(1-e^{-\tau}),
\end{equation}
where $\tau$ is the Lyman continuum optical depth through a given cloud, and $\partial^2 N/(\partial N_{HI} \partial z)$ is the absorber distribution that we specified for our model in Eq. \ref{eq:lyforest}. For non-ionizing radiation, the effective opacity is zero.  At wavelengths well above the Lyman limit, the background at a redshift $z_{0}$ is determined by the total history of emission at higher redshifts $z>z_{0}$.  At ionizing wavelengths, the mean free path of photons is shorter than cosmological distances at redshifts greater than the breakthrough redshift, and therefore the ionizing background at $z_{0}>z_{br}$ is determined by the emissivity of approximately contemporary sources.  

In Fig. \ref{fig:eblfluxUV} we present the $z=0$ EBL flux for our semi-analytic models, alongside two other recent EBL determinations using alternate methods.  At low redshift, the effect of our high quasar and high-peaked star-formation models on the total background are negligible, and are not shown.  We have shown the EBL calculated out to the far-IR in this plot, though results at wavelengths longer than the optical--near-IR peak at $\sim 1 \mu$m are not relevant to gamma-ray opacities in the high-redshift regime we discuss here.  Dust reemission of light in this model is based upon the IR templates of \citet{devriendt00}.  Much more detailed IR background modeling, taking into account the latest {\it Spitzer} data, will be presented in \citet{gilmoreEBL}.    

Understanding the evolution of the background
field is critical to our calculation of the absorption of gamma rays,
as the flux at high redshift can be both significantly higher than and
of a different spectral shape than the local background.  In
Fig.~\ref{fig:eblhistUV} we show the background flux for our four
models at several redshifts, including $z=0$ where we have also shown
a compilation of observable data, including estimates from both
absolute photometry and discrete source number counts.   In each case, we assume the `best-fitting' escape fractions of Table
\ref{tab:models}, which we have chosen based on the comparisons with
Ly$\alpha$ forest measurements.  At all redshifts, the background
shows a sharp drop at the Lyman edge; this is a combined consequence
of absorption in stellar atmospheres, H I in galaxies (quantified as
\feschi\ in our models), and IGM reprocessing.  The feature at $\sim$300 \AA~is
due to He II Ly$\alpha$. 

\section{Gamma-ray Attenuation}

Gamma rays can interact with background photons if sufficient energy
exists in the centre of mass to create an electron-positron pair with total mass $2m_e$:
\begin{equation}
\sqrt{2E_1 E_2 (1-\cos\theta)} \geq 2 m_e c^2,
\end{equation}
where $E_1$ and $E_2$ are the photon energies and $\theta$ is the
angle of incidence.  A gamma-ray of energy $E_\gamma$ can therefore pair-produce with background photons above a threshold energy of
\begin{equation}
E_{th}=\frac{2m_e^2c^4}{E_\gamma (1-\cos\theta)}
\end{equation}
The cross-section for this process is
\citep{madau&phinney96}
\begin{eqnarray}
\lefteqn{\sigma(E_1,E_2,\theta) = \frac{3\sigma_T}{16}(1-\beta^2)} \nonumber \\
& & \times \left[ 2\beta(\beta^2-2)+ (3-\beta^4)\ln \left( \frac{1+\beta}{1-\beta}\right)\right], 
\end{eqnarray}
where
\begin{equation}
\beta = \sqrt{1-\frac{2m_e^2c^4}{E_1 E_2 (1-\cos\theta)}},
\end{equation}
and $\sigma_T$ is the Thompson scattering cross section.  The cross
section is maximized for centre of mass energies of approximately
twice the threshold energy $2m_e c^2$, and falls as approximately as inverse energy for
$E \gg E_{th}$.  The likelihood of absorption is maximized for photons
at about 4 times the absolute threshold energy, with one factor of 2
from $\sigma$ and another in going from $\theta=\pi$ (`head-on'
configuration) to the most probable angle of interaction $\theta
\approx \pi /2$.  Gamma rays above 1 TeV are most attenuated by the
near- and mid-IR range of the EBL, while those in the 300 GeV to 1 TeV
regime are sensitive to light in the near-IR and optical. Below 200
GeV it is mainly UV photons that have sufficient energy to cause the
pair-production interaction.  Below 19 GeV only background photons with energies above the
Lyman limit have sufficient energy to interact at any angle in the
rest frame, and there is little attenuation.  Note that these numbers refer to rest-frame energy of the gamma ray, which can be substantially higher than its observed energy due to redshifting.   If the differential number density of
background photons at energy $E_{bg}$ is $n(E_{bg},z)$, then the
optical depth of attenuation for a photon of observed energy $E_\gamma$ is
\begin{eqnarray}
\lefteqn{\tau(E_\gamma,z_0) =  \frac{1}{2}\int^{z_0}_0 dz\;\frac{dl}{dz}\int^1_{-1}d(\cos\theta) \; (1-\cos\theta)} \nonumber \\ 
& & \times \int^{\infty}_{E_{min}} dE_{bg}\; n(E_{bg},z)\;\sigma(E_\gamma (1+z),E_{bg},\theta).
\label{eq:opdep}
\end{eqnarray}
\noindent Where we have \[E_{min}=E_{th}\:(1+z)^{-1}=\frac{2m_e^2c^4}{E_\gamma (1+z)(1-\cos\theta)}\] and $dl/dz$ is the cosmological line element defined in Equation \ref{eq:cosline}.

For each of our four models, we have calculated the optical depth of
gamma rays at all relevant energies and redshifts.  As in our
calculation of the background flux above, we assume the H I escape
fractions listed in Table \ref{tab:models}.  It should be emphasized
that the choice of escape fraction has little effect on absorption of
gamma rays at energies $>10$ GeV.  We find, as argued in \citet{oh01},
that the background field at energies above 1 Ry is negligible as a
barrier to cosmological gamma rays, and that significant optical depth
above this energy is due to photons longward of the Lyman limit, where
photon density increases dramatically in all of our models.  While
gamma rays are limited to interactions with background photons of an
absolute minimum energy $E_{th}=m_e^2c^4/E_\gamma$ (with
$\cos\theta=-1$), redshifting places these gamma rays at higher
energies at earlier epochs, where they can pair produce on the
non-ionizing background.  The increase of star-formation rate density
by roughly an order of magnitude between present-day and peak rates
means that gamma rays from high-redshift sources will tend to be
attenuated most strongly at these early redshifts, where they have
energies ($1+z$) times higher than at $z=0$.

\begin{figure*}
\psfig{file=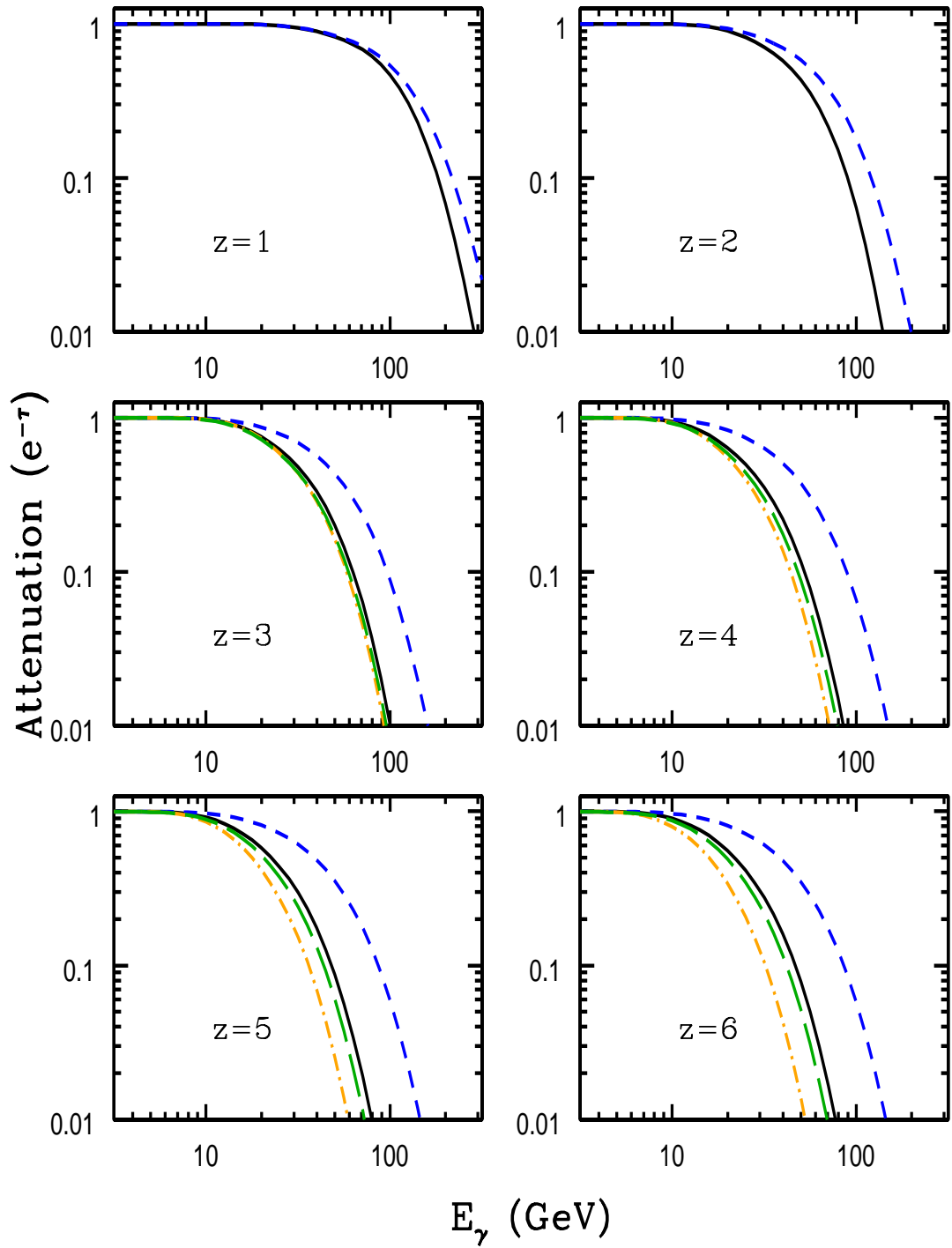,width=15cm,height=20cm}
\caption{Attenuation factors ($e^{-\tau}$) as a function of gamma-ray
  energy for the indicated source redshifts.  Curves are as in
  Fig.~\ref{fig:eblhistUV}, and indicate the absorption resulting from
  our models of star-formation and quasar emissivity.  Black solid
  line: fiducial model with H I escape fraction of 0.1.  Dashed blue
  line: low star-formation model, with escape fraction 0.2.  Orange
  dash-dotted line: high-peaked star formation rate with escape
  fraction 0.1.  Green long-dashed: fiducial model with SB03 quasar
  contribution and escape fraction 0.02.  Curves for the high-peaked
  star formation and high quasar models converge to the fiducial model
  for $z \leq 2$.}
\label{fig:opdep_multi}
\end{figure*}

In Fig.~\ref{fig:opdep_multi}, the optical depth vs gamma-ray energy is
shown for each model at various redshifts.  These high-redshift
results should be considered complementary to the our other
calculations of EBL with these semi-analytic models, which emphasized
the absorption of $>100$ GeV gamma rays at lower redshift.  The effect
of the UV background is to produce a relatively sharp and featureless
cutoff in energy.  At energies above 100 GeV, the effect of the EBL
has often been quantified as a change in the spectral index of
observed blazar spectra \citep[e.g.][]{aharonian06}, due to the
relatively flat number density of EBL photons in the near and mid-IR.
At lower energies, this approximation is not valid over any
significant range in energy, due to the steepness of the cutoff that
results in rapidly increasing numbers of photons with increasing
wavelength in the UV.


Our high-peaked SFRD and quasar-dominated models give absorption
features that are similar, despite being very different in terms of
the spectral form of the background flux.  While the emission from
quasars produces a much higher ionizing background, the spectral
cutoff at all redshifts we have explored is dominated by the photons
longward of the Lyman limit.
     
\begin{figure}
\psfig{file=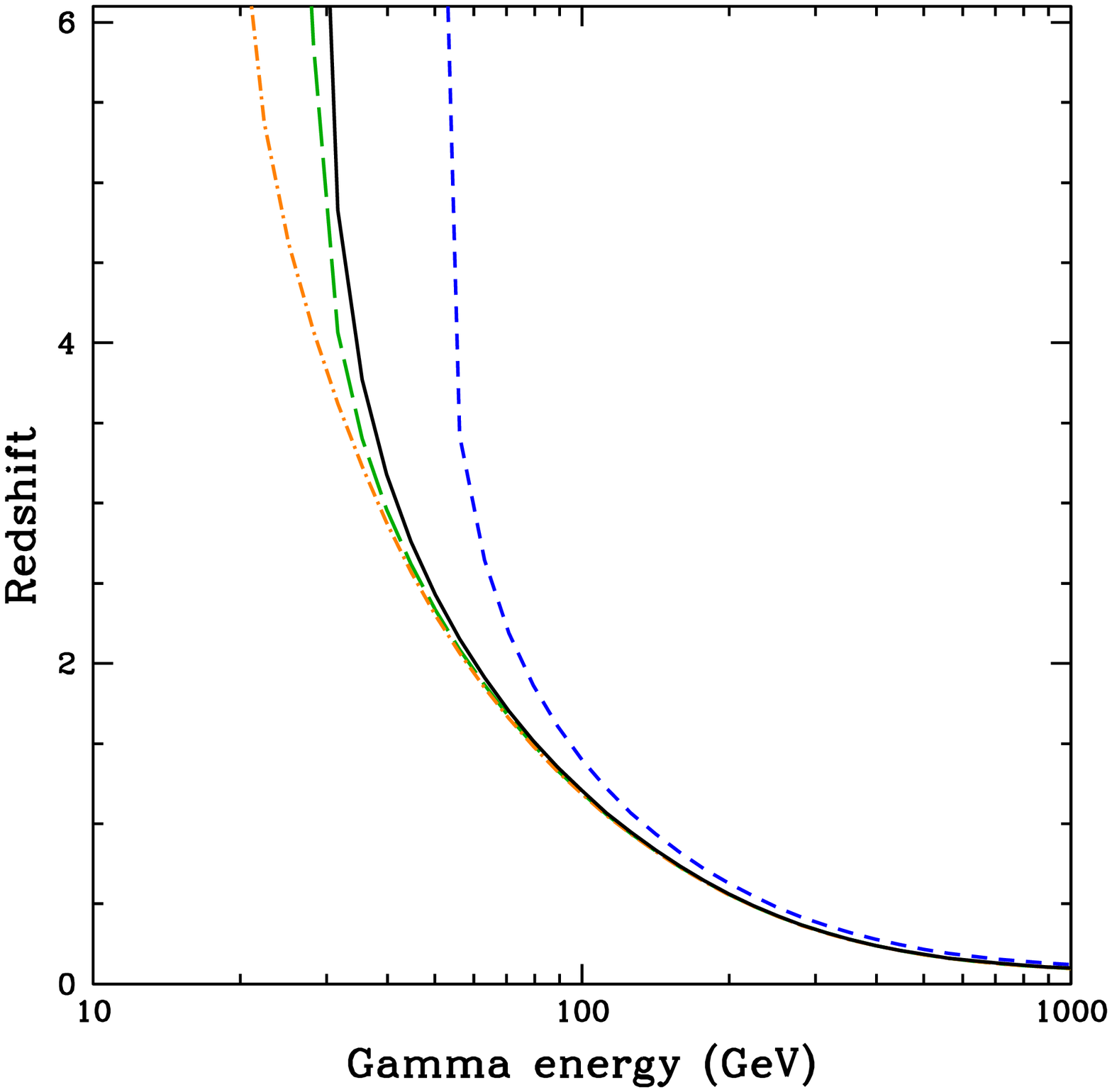,width=\columnwidth}
\caption{The redshifts at which the universe becomes optically thick ($\tau>1$) to gamma rays at a given observed energy.  Line colours and types are as in Fig. \ref{fig:opdep_multi}. }
\label{fig:attedgeUV}
\end{figure}

In Fig.~\ref{fig:attedgeUV}, we show the redshifts at which the universe becomes optically thick ($\tau>1$) to gamma rays of a given energy for each of our models.  From this plot, we can see in a general sense how background attenuation affects different energy regimes at different redshifts.  The low model shows little change at redshifts higher than about 3, due to the rapid decline in star formation after this point.  The high-peaked model has the most impact at high redshift, and produces absorption features that evolve out to $z>6$. 

\section{Discussion}

\begin{table*}
\caption{Here we summarize a broad set of models, and the qualitative
  level of agreement of each with Ly$\alpha$ forest data, proximity
  effect measurements, and simulations of total ionizing escape
  fraction from star-forming galaxies.  The numbers of the models
  shown in Figs.~ \ref{fig:eblhistUV}, \ref{fig:opdep_multi}, and \ref{fig:attedgeUV} are in
  bold.}

 \label{tab:modelsum}
 \begin{tabular}{|llll|cccc|}
\hline & \hspace{1cm}{\bf Model Parameters} & & & & {\bf Fits with
  Data} & & \\ \hline Model & SFR Density & Quasar Luminosity &
$f_{escHI}$ & Flux Decrement$^a$ & Proximity Effect$^b$ & Softness$^c$
& $f_{esc}$$^d$ \\ \hline \hline 1.1 & Fiducial & HRH07 & 0.02 & x & x
& $\circ$ & $\surd$ \\ \hline {\bf 1.2} & Fiducial & HRH07 & 0.10 &
$\circ$ & x & $\surd$ & $\surd$ \\ \hline 1.3 & Fiducial & HRH07 &
0.20 & $\circ$ & x & $\circ$ & $\surd$ \\ \hline 2.1 & Low & HRH07 &
0.10 & x & x & $\circ$ & $\surd$ \\ \hline {\bf 2.2} & Low & HRH07 &
0.20 & $\circ$ & x & $\surd$ & $\surd$ \\ \hline 2.3 & Low & HRH07 &
0.50 & $\circ$ & x & $\circ$ & $\circ$ \\ \hline {\bf 3} & Fiducial
High-peaked & HRH07 & 0.10 & $\surd$ & x & $\surd$ & $\surd$ \\ \hline
{\bf 4} & Fiducial & SB03 Model C & 0.02 & x & $\surd$ & x & $\surd$
\\ \hline
\end{tabular}

\medskip
\begin{flushleft}
$\surd$ -- best agreement \\
$\circ$ -- marginal agreement \\ 
x -- poor agreement \\

\medskip
$^a$ This refers to the relatively flat ionization rate for H I at a level seen in the quasar spectra data of \citet{bolton05} and \citet{faucher08} (Fig.~ \ref{fig:ionrates}). \\
$^b$ The higher levels of ionization rate determined in uncorrected proximity effect measurements (Fig.~ \ref{fig:ionrates}).\\
$^c$ The softness $\eta \equiv N({\rm HeII})/N(\rm{HI})$, from data compiled in Fig.~ \ref{fig:heii_hi}. \\
$^d$ The total escape fraction; recall from Section 2 that this is equivalent to $\feschi \ast f_{1500}$.  In our semi-analytic models, $f^{-1}_{1500}$ ranges from about 7 to 14 at the redshifts of interest.
\end{flushleft}
\end{table*}

We have created and analysed predictions for the UV background that
are intended to broadly span the possibilities in star-formation rate
and quasar-luminosity density.  Our fiducial model with the HRH07 QSO
LF (first entry in Table \ref{tab:models}) provides a reasonable match
to the level of ionizing flux inferred from Ly$\alpha$ forest
measurements when an H I escape fraction of 0.1 is assumed; this
corresponds to a total $f_{\rm esc}$ of 1 to 1.5 per cent when
combined with the dust absorption values predicted by our
semi-analytic galaxy formation model.  The `low' model for the SFR
density, with the lower CDM power spectrum normalization of WMAP3,
requires a larger escape fraction to match ionization rate data,
especially the higher redshift points $z>5$, where $\feschi \sim 0.5$
is required.  Both of these models fail reproduce the nearly constant
ionization rate between $2<z<4.5$ seen in some flux decrement
analyses.  The high-peaked model, which has a star-formation rate that
increases until redshift 5, does produce a somewhat flatter ionization
rate, as suggested by \citet{faucher08a}.  The large amount of high-redshift star-formation in this model is not supported by estimates of stellar mass buildup, and should be considered a somewhat extreme scenario.  Another mechanism for producing a flatter ionization rate history is an evolving escape fraction that increases with
redshift, a possibility that we do not explore here but which has been
seen in some simulations \citep{razoumov&sommerlarsen07,razoumov&sommerlarsen06}, and as
already mentioned may be suggested by observations which have detected
Lyman continuum radiation from distant ($z \sim 3$) galaxies
\citep{shapley06}, but not closer sources \citep{siana07}.  This would have only a weak effect on the opacities we have calculated, as most attenuation of gamma rays is due to the non-ionizing UV background, which would have a much larger number density than the ionizing background even for a high escape fraction.  

Our results suggest that observations of sufficient numbers of
high-energy gamma-ray sources out to high redshift could provide a
probe of the UV background at these epochs that is independent of any
other observational test.  Pair-production with target background
photons produces a spectral cutoff at energies that are dependent upon
redshift and assumed cosmological model.  With enough detections of
blazars and/or gamma-ray bursts (GRBs) at different confirmed
redshifts in the 10 to 100 GeV energy decade, it should be possible to
detect an evolving cutoff in energy, and distinguish between the
different background levels proposed in this work.  The exact number
of blazars that will be detected at GeV energies over the coming years
is uncertain and dependent upon the poorly-understood emission
processes and number density evolution of these sources.  However,
even conservative estimates indicate that a large number of sources
will be detectable by the {\it Fermi}~spacecraft, which was launched
in June 2008 and is currently in all-sky survey mode.  The EGRET
experiment on the {\it Compton Gamma-ray Observatory} ({\it CGRO})
detected more than 60 high-confidence blazars at energies of $>100$
MeV out to redshift 2.28, mostly of the flat-spectrum radio quasar
(FSRQ) type \citep{mukherjee97}.  An extrapolation of these results
suggests that {\it Fermi} will see $\sim 1000$ blazars extending to
higher redshift \citep{dermer07}.  An analysis of two different
realizations of the blazar luminosity function by \citet*{chen04}
suggested that {\it Fermi} could detect thousands of blazars, and
would potentially be able to measure attenuation at distances as great
as $z=5$.  The 3-month {\it Fermi} LAT survey has already reported 106 AGN sources at high confidence \citep{abdo09a}.  In addition to analyzing blazar spectra in survey and
pointed observations, {\it Fermi} will also act as a finder for new
and upcoming ground-based experiments such as H.E.S.S.-II and MAGIC-II
which will be capable of resolving most of the energy ranges of
interest.  In survey mode, {\it Fermi} will also act as an alert
system for flaring sources.

None of our models predicts significant attenuation at 10 GeV or below
for any redshift.  This is true even for our extremely
quasar-dominated model, where the opacity of a 10 GeV observed gamma
ray is never higher than $\tau \sim 0.2$.  As the ionizing flux in
this model is higher than allowed by most measurements of the
Ly$\alpha$ forest, it is unlikely that any cosmological model could
produce significant gamma-ray opacity due to a large contribution of
ionizing photons to the background.  The high-peaked star formation model produces the most absorption in the 10--100 GeV energy range for $z>3$, but despite having a very high UV output only has a moderate impact on the calculated optical depths relative to the fiducial model. 

\subsection{Comparison with other work}

It is useful to compare the absorption predicted by our models with
the calculations of other authors who have used different methods, in
the cases where their results include our energy and redshift
regime of interest.  In many instances, our predicted attenuation is
less than what has been previously proposed.  

The background model of
\citet{franceschini08} is based upon extrapolated luminosity functions determined from a large compilation of
multiwavelength data, including deep ACS imaging of distant galaxies, and treats
separately the evolutionary histories of spiral, elliptical, and star-bursting galaxy populations.  While their EBL agrees well with our fiducial model at z=0 and z=1
\citep{gilmoreEBL}, their absorption $\tau$ in the 10--100 GeV energy
decade is at least a factor of two greater at $z=$2--4 than any of our
models.  The most recent models of Stecker and collaborators
\citep{stecker06,stecker07} are based on a
`backwards evolution' model in which galaxies' emission SEDs are
determined by their brightness in one band, taken to be 60 $\mu m$.  The luminosity of the galaxy population at this wavelength is assumed
to brighten with redshift as a power law in ($1+z$).  One disadvantage
of this method is that it attempts to describe luminosity evolution
over several orders of magnitude in wavelength from a single broken power
law, which cannot take into account the complexity of galaxy
evolution.  Gamma-ray opacities in this work are much higher than our
predictions, with the universe optically thick ($\tau>1$) to 10 GeV
gamma rays above $z\sim3$, and for $>25$ GeV above $z=1$.  This level
of absorption holds very different implications for experiments such
as {\it Fermi}.  At high redshifts, absorption cutoff spectral
features would be visible between about 5 and 20 GeV, with no signal
from higher energies due to optical thickness from the background.
The galaxy SEDs in these models have no emission above Lyman energies,
and therefore all attenuation at these very low energies is the result
of near-threshold interactions with non-ionizing UV photons.  The redshift-dependent optical and UV SEDs used are based on the population synthesis models of \citet{bruzual&charlot93}.  This model does not include UV dust extinction, which as we have found in this work can reduce far-UV emissivity by a factor of $\sim$10 at higher redshifts.

The recent observation by {\it Fermi} of high-energy emission from GRB
080916C at z=4.35 \citep{greiner09} provides a valuable first test of
these predictions for GeV absorption.  The highest energy photon seen
by the LAT was 13.2 GeV, with over 10 photons seen above 1 GeV
\citep{abdo09a}.  In all of our models, the gamma-ray optical depth is
much less than 1 for this energy and redshift, and similar values are
found in the star-formation models of \citet{razzaque09}.  The models
of Stecker and collaborators predict a much higher opacity, $\tau=3.5$
to 4.5, for the 13 GeV photon.  Including the 1$\sigma$ error on redshift and photon energy in finding maximal and minimal values, the corresponding transmission probability could be as high as 8.2 per cent for their `baseline' model, or as low as 0.5 per cent for the `fast evolution' model (Stecker, private communication).   While it is difficult to draw
conclusions from a single event, more bursts seen with GeV emission
equal or greater to GRB 080916C could strongly disfavor such a large
background flux, and observations of slightly higher energy photons
($E \sim 30$ GeV) from similar redshifts could provide a test of our
models.

\subsection{Caveats and future work}

We expect our approach to be reasonably accurate at predicting the
ionizing and non-ionizing background fields out to redshift $\sim 6$,
where H I Gunn-Peterson troughs appear in observed quasar spectra
\citep{fan06}.  At higher redshifts, during the epoch of reionization,
the concept of a uniform background for ionizing photons is no longer
valid, as photons above the Lyman limit are confined to the vicinity
of their sources.  In our Lyman absorption model (Equation
\ref{eq:lyforest}), the sudden increase seen in H I opacity at
redshift six is not represented, and our model would therefore be
expected to overproduce the ionizing background above this redshift. A
similar limitation exists in our treatment of He opacities above the
redshift of He reionization $z\sim 3$.  However, these factors alone
are unlikely to have a significant effect on calculated opacities.

We have made the assumption of a universal stellar IMF in this work,
and have not included a separate population of metal-free
(population-III) stars or other early source types such as miniquasars
\citep{madau04}.  These types of unobserved sources could have very
different spectra than standard stellar populations, and could produce
large contributions to the ionizing and non-ionizing UV backgrounds.
Because of redshift effects, gamma rays with low ($<10$ GeV) observed
energies for very high-redshift sources could have significant
interactions with the freely-propagating non-ionizing background.  It
is therefore possible that opacities at reionization redshifts could
be much higher than we propose here due to unseen UV production
mechanisms.  While models for gamma-ray blazars do not typically
predict sources at these high redshifts, GRBs are known to exist above
redshift six \citep{greiner09a}, and long-duration GRBs could
potentially be seen as far out as star-formation occurs.  The EGRET
experiment, operating from 30 MeV up to $\sim$30 GeV, was able to view
a small number of photons from GRBs, and the detection of high-energy
emission from GRB 080916C by the {\it Fermi} LAT demonstrates the
ability of this instrument to detect GeV photons from these events.
Though predictions are highly uncertain, it is possible that GRBs
could produce significant numbers of photons well above 10 GeV through
inverse-Compton or hadronic processes \citep*{le&dermer08,ando08}.
Calculations of the background flux from some possible reionization
scenarios and source types at $z>6$ may therefore be a worthwhile
undertaking.

It has been suggested by a number of authors that the discrepancy
between observed stellar mass density and instantaneous star-formation
rate density (see Section~\ref{sec:sfr}) could be explained by an IMF
that evolves with redshift or is more top-heavy in rapidly
star-forming galaxies \citep{dave08,fardal07,baugh05}.  Alternatively,
an IMF with shallower high-end slope has been suggested as a source of
early reionization \citep{chary08}.  Altering the high-mass end of the
IMF will change the spectrum produced by galaxies and also the
attenuation by dust, although as probes of star-formation generally
involve the same high-mass stars that produce the UV background, there
is some degree of degeneracy between these two quantities when the
assumed IMF is changed.  This is also an issue that warrants further
study.

\section*{Acknowledgments}
R.G. and J.P. acknowledge support from a Fermi Guest Investigator
Grant and NSF-AST-0607712, and J.P. also acknowledges support from a
NASA ATP grant. Support for this work was provided to P.M. by NASA
through grants HST-AR-11268.01-A1 and NNX08AV68G.

\bibliographystyle{mn2e}

\label{lastpage}
\end{document}